\documentclass{aa}  
\usepackage[varg]{txfonts}
\usepackage{graphicx}
\usepackage{txfonts}
\usepackage{amstext}
\usepackage{xcolor}
\usepackage{color}
\usepackage{threeparttable}
\usepackage{overpic}
\usepackage{multirow}
\usepackage[normalem]{ulem} 
\usepackage{amsmath} 
\usepackage{soul}
\usepackage[normalem]{ulem}
\usepackage{hyperref}
\usepackage{orcidlink}
\hypersetup{
  colorlinks   = true, 
  urlcolor     = blue, 
  linkcolor    = blue, 
  citecolor   = blue 
}
\usepackage{placeins}

\newcommand{\ms}{M$_{\sun}$}

\newcommand{\ond}{Ond\v{r}ejov}
\newcommand{\obj}{EI~Psc }
\newcommand{\T}{{\sc TESS}}

\begin{document} 

   \title{Hot donors in cataclysmic variables: The case of EI~Psc}

   \author{J. K\'{a}ra \inst{1, 2,}\orcidlink{0000-0002-1012-7203}\thanks{Corresponding 
   author: \href{mailto:honza.kara.7@gmail.com}{honza.kara.7@gmail.com}}
      \and S. Zharikov\inst{3, 4}\orcidlink{0000-0003-2526-2683}
      \and M. Wolf~\inst{1}\orcidlink{0000-0002-4387-6358}
      \and N. Vaidman\inst{4, 5}\orcidlink{0000-0002-7449-0108}
      \and A. Agishev\inst{4}\orcidlink{0000-0001-9788-7485}
      \and S. Khokhlov\inst{4}\orcidlink{0000-0001-5163-508X}
      \and C.~E. Chavez\inst{6}\orcidlink{0000-0002-6900-1154}
        }
          
\institute{Astronomical Institute, Faculty of Mathematics and Physics, Charles University,
V~Hole\v{s}ovi\v{c}k\'ach~2, CZ-180~00~Praha~8, \\
            Czech Republic
\and Department of Physics and Astronomy, University of Texas Rio Grande Valley, 
Brownsville, TX 78520, USA
\and Universidad Nacional Aut\'{o}noma de M\'{e}xico, Instituto de Astronom\'{i}a, AP 106,  
Ensenada, 22800, BC, M\'{e}xico 
\and Faculty of Physics and Technology, Al-Farabi Kazakh National University, Al-Farabi 
Ave., 71, Almaty 050040, Kazakhstan
\and Fesenkov Astrophysical Institute, 23 Observatory Str., Almaty 050020, Kazakhstan  
\and Facultad de Ingenier\'{i}a Mec\'{a}nica y El\'{e}ctrica, Universidad Aut\'{o}noma de 
Nuevo Le\'{o}n, Monterrey, Nuevo Le\'{o}n, 66451, M\'{e}xico}

 \date{Received \today}
 
  \abstract
   {We present results of time-resolved optical spectroscopy and photometry
   of the short-orbital period dwarf nova \object{EI Psc}.}
   {This study aims to determine fundamental system parameters of EI~Psc, study 
   properties of accretion structures in the system, and investigate its origin and current 
   evolution state.} 
   {We analyse newly obtained time-resolved spectroscopic and photometric observations as 
   well as archival data. We used light curve modelling, Doppler tomography, and {\sc mesa} 
   evolutionary models to study the characteristics of EI~Psc. 
   }
   {The system contains a relatively low temperature ($T_{\rm eff} = 6130$ K) white dwarf 
   with  mass of $M_{\mathrm WD} = 0.70(4)$~\ms. The mass of the warm  ($T_2 = 4440$~K) 
   secondary is $M_2 = 0.13$~\ms. The inclination of the system is $i= 44\fdg5(7)$. The 
   mass accretion rate is  $\approx$ 4$\times$10$^{-13}$ M$_\odot$ year$^{-1}$.  
   The long-term light curve of the system shows outbursts and superoutbursts. The 
   quiescence light curve is double-humped and is formed by the combination of radiation 
   from the Roche lobe filling the hot secondary and the hot spot. The radius of the outer 
   disc is about two times smaller than the tidal truncation radius. 
   Most of the disc's emission consists of emission lines and radiation from the hot spot 
   at the stream-disc impact region.
   }
    {These types of systems are formed from progenitors with a low mass 
    WD M$_{\mathrm{WD}} \lesssim 0.6$ M$_\odot$ and relatively massive 
    secondaries $1.1-1.5$ M$_\odot$ with 
    initial orbital periods on a scale of days. The number of similar systems is expected 
    to be significantly lower than the usual CVs due to a lower forming rate of their 
    relatively massive progenitors.}


   \maketitle

\section{Introduction}
\label{Intro}

Cataclysmic variables (CVs) are close-interacting binary systems 
consisting of a white dwarf (WD) as a primary and a low-mass 
main-sequence star or a brown dwarf as a secondary component.
The Roche-lobe filling secondary component loses matter through the 
inner Lagrangian point to the primary \citep{1995cvs..book.....W}. 
In the absence of a strong magnetic field, the material transferred 
from the donor forms an accretion disc around the WD and eventually 
accretes. According to standard evolutionary theory, CVs evolve from
longer orbital periods towards shorter ones down to a period minimum 
($\approx$70-80~min depending on orbital angular momentum losses)
when the secondary is driven out of thermal equilibrium and becomes 
partially degenerate \citep{1981ApJ...248L..27P, 1982ApJ...254..616R, 
2011ApJS..194...28K}. Short-period CVs tend to have low-mass 
secondaries ($<0.1 M_\odot$), which are close to the expected 
main-sequence star surface temperature.  However, there are exceptions. 
A small group of systems has anomalously warm donors, typically of the 
K spectral type: \object{EI Psc }($P_{\rm orb}=64$ min, K4(2), 
\citet{2002ApJ...567L..49T, 2016ApJ...833...14H}), 
\object{QZ Ser} ($P_{\rm orb}$=120 min, K4(2) \citet{2002PASP..114.1117T}),
\object{SDSS J170213.26+322954.1} ($P_{\rm orb}$=144 min, M0, \citet{2006MNRAS.371.1435L}), 
\object{CRTS J134052.1+151341} ($P_{\rm orb}$=147 min, K4(2), \citet{2013PASP..125..506T}),  
\object{ASASSN-18aan} (215 min, G9, \citet{2021PASJ...73.1209W}), and
\object{ASAS-SN 13cl} (292 min, K4(2) \citet{2015PASP..127..351T}). 
Possible evolutionary scenarios for these systems were briefly explored by 
\citet{2002ApJ...567L..49T}, who suggested that the secondary 
is substantially enriched in helium due to nuclear evolution 
before onset of the mass transfer between components. However, 
the evolutionary status and the driving mechanism of the mass 
transfer of these objects are open issues.

The variable star \object{EI Psc} (also \object{1RXS J232953.9+062814}, 
\object{2MASS J23295420+0628116}, \object{TIC 423324616},  
\object{Gaia DR3 2757417008882236672}) was discovered as an X-ray 
source by the ROSAT satellite and identified as a cataclysmic variable by
\cite{1998AnShO..19..235H}.  \cite{2001IAUC.7747....2U} reported that 
their photometric observations showed superhumps with amplitudes 
ranging from $0.2$ to $0.3$ mag and period 
$P_{\mathrm{sh}}=0.046311(12)\,\mathrm{d}$. 
\cite{2002ApJ...567L..49T} used spectroscopic observations to derive 
the orbital period $P_{\mathrm{orb}}=0.044566(3)\,\mathrm{d}$ and 
estimated the mass ratio to be $q=M_2/M_{\rm WD}=0.185$  based on the 
superhump period excess $\epsilon = (P_{\rm sh}-P_{\rm orb})/P_{\rm orb}$ 
using an $\epsilon\mathrm{-}q$ relation derived by 
\cite{1998PASP..110.1132P}. 
The orbital period of \object{EI~Psc} is below the period minimum of CVs with
main-sequence secondaries. Usually, such periods correspond to 
AM~CVn stars; helium CVs with absent hydrogen emission in their spectra,
which are considered to have a secondary star of a helium white dwarf 
or a semi-degenerate helium star \citep{2005ASPC..330...27N}.   
However, the low-resolution quiescent spectra presented by 
\cite{2001ChJAA...1..483W} revealed that \object{EI~Psc} is a hydrogen-rich system.
The spectra analysed by \cite{2002ApJ...567L..49T} showed a prominent 
broad Balmer emission, along with weaker emission from \ion{He}{I}. 
Balmer lines exhibited double peaks for the majority of the time and 
were separated by $1000 \pm100$~km~s$^{-1}$ when the peaks were most 
distinct. The spectrum also showed absorption features that appeared 
to be consistent with a K4-type star in the spectral library from 
\citep{1998PASP..110..863P}. The absorption spectrum exhibited an 
obvious modulation and the $H_\alpha$ emission line showed an apparent 
S-wave. EI~Psc  was also observed in the far ultraviolet (FUV) using 
HST/STIS by \citet{2003ApJ...594..443G}. The authors found that the 
FUV continuum of EI~Psc is extremely red, suggesting that the white
dwarf in this system is very cold. They also found that the object 
displays anomalously high \ion{N}{v}/\ion{C}{iv} emission-line flux 
ratios. Such anomalous line flux ratios are expected in CVs that 
went through a phase of thermal timescale mass transfer and now 
accrete CNO-processed material from a companion stripped of its external 
layers. Infrared spectroscopy was obtained by \citet{2009AJ....137.4061H, 
2016ApJ...833...14H}. They found that the spectral type appears to be
close to K5V. However, they note that there is an issue with this 
spectral classification because the \ion{Ca}{i} triplet is  much weaker
than the \ion{Na}{I} doublet, in contrast to typically K dwarf spectra, 
which show a similar strength of those spectral features. In addition, 
no evidence of CO absorption features was reported. 
\citet{2009AJ....137.4061H} also modelled the g-band and the JHK infrared
light curves of the object and estimated parameters of the secondary as 
$M_2$ = 0.13 M$_\sun$, $R_2$ = 0.12 R$_\sun$, the system inclination of 
$i=55\degr$, and the distance to the system of about 150~pc.
The {\sc Gaia} parallax from DR3 of 6.62(5)$\times 10^{-3}$ arcsec 
corresponds to the distance of 148.5(1.0)~pc \citep{2018yCat.1345....0G, 
2021AJ....161..147B}  and it confirms estimation of 
\cite{2009AJ....137.4061H}. 

Here, we present new time-resolved photometric and spectroscopic
observations of the object to study its nature, origin, and structure of 
the accretion flow in such an unusual CV. The paper is structured as 
follows. In Sections \ref{DataObsRed},~\ref{s:lightCurve}, and 
\ref{S:Spectra}, we describe our photometric and spectroscopic 
observations and general data analysis. Section~\ref{S:LCM} presents 
the results of the modelling of the quiescence light curve, while
Doppler tomography is discussed in Section~\ref{S:DT}. The evolution 
state of the system is explored in Section~\ref{S:Evol} and our 
conclusions are summarised in Section~\ref{S:Conclus}.

\section{Observation and data reduction}
\label{DataObsRed}

\begin{table*}[htb]
    \caption{Log of spectroscopic observations of EI~Psc.}
    \centering
    \begin{tabular}{l c c c @{ -- } l  c c c}
\hline\noalign{\smallskip}															
Instrument	&	Date	&	JD	&	\multicolumn{2}{c}{Spectral range}			&	$R =  \lambda / \Delta \lambda $	&	Exposure	&	Number of 	\\
	&		&	$-2\,400\,000$	&	\multicolumn{2}{c}{[\AA]}	&	& time	[s]	&	exposures	\\
\hline\hline\noalign{\smallskip}	
LAMOST	&	2013, Sep. 25	&$	56\,561	$&$	3700	$&$	9000	$&$	1800	$&	4500 & 1 \\
\hline\noalign{\smallskip}															
VLT (X-Shooter)	&	2021, Sep. 7	&$	59\,464	$&$	2989	$&$	5560	$&$	5453	$&	436	&	3	\\
	&	&		&$	5337	$&$	10\,200	$&$	8935	$&	420	&	3	\\
	&	&		&$	9940	$&$	24\,790	$&$	5573	$&	480	&	3	\\
\hline\noalign{\smallskip}
GTC (Osiris+)	&	2023, Sep. 4	&$	60\,192	$&$	7330	$&$	10\,000	$&$	2503	$&	181	&	29	\\
	&	2023, Sep. 5	&$	60\,193	$&$	5575	$&$	7685	$&$	2475	$&	181	&	29	\\
	&	2023, Sep. 7	&$	60\,195	$&$	4500	$&$	6000	$&$	2515	$&	181	&	29	\\
\hline\noalign{\smallskip}																							
    \end{tabular}
    \label{T:LOG:SP}
\end{table*}

\begin{table}
    \caption{Log of photometric observations of EI~Psc.}
    \centering
    \begin{tabular}{l c c c c}
\hline\noalign{\smallskip}											
Date	&	JD	            & Band/  	&	Exp.   &	Dur.	\\
        & $-2\,400\,000$	& Sector	&	[s]	   &		\\
\hline\hline\noalign{\smallskip}										
\multicolumn{5}{c}{ \underline{\makebox[0.46\textwidth][c]{SPM}} } \\
		2023, Nov. 2	&$	60\,250	$&	V	&	60	&	4.92 h	\\
		2023, Nov. 3	&$	60\,251	$&	V	&	60	&	2.95 h	\\
		2023, Nov. 4	&$	60\,252	$&	V	&	60	&	4.83 h	\\
		2023, Nov. 5	&$	60\,253	$&	V	&	60	&	3.90 h	\\
		2023, Nov. 6	&$	60\,254	$&	V	&	60	&	3.60 h	\\
		2023, Nov. 7	&$	60\,255	$&	V	&	60	&	4.97 h	\\
		2023, Nov. 8	&$	60\,256	$&	V	&	60	&	5.80 h	\\
		2023, Nov. 24	&$	60\,272	$&	V	&	60	&	3.40 h	\\
\hline\noalign{\smallskip}											
\multicolumn{5}{c}{	\underline{\makebox[0.46\textwidth][c]{\ond}}		}								\\
		2023, Aug. 10	&$	60\,167	$&	C	&	60	&	0.38 h	\\
		2023, Aug. 11	&$	60\,168	$&	C	&	60	&	0.79 h	\\
		2023, Sep. 20	&$	60\,208	$&	C	&	60	&	3.27 h	\\
		2023, Sep. 26	&$	60\,214	$&	R	&	60	&	4.53 h	\\
		2023, Oct. 17	&$	60\,235	$&	V	&	60	&	2.34 h	\\
		2024, Oct. 24	&$	60\,608	$&	C	&	60	&	0.82 h	\\
		2024, Oct. 28	&$	60\,612	$&	C	&	60	&	1.22 h	\\
		2024, Nov. 1	&$	60\,616	$&	C	&	60	&	1.54 h	\\
		2024, Nov. 8	&$	60\,623	$&	C	&	60	&	0.87 h	\\
		2024, Nov. 9	&$	60\,624	$&	C	&	60	&	1.29 h	\\
        2024, Nov. 25	&$	60\,640	$&	C	&	60	&	1.17 h	\\
        2024, Nov. 27	&$	60\,642	$&	C	&	60	&	1.11 h	\\        
\hline\noalign{\smallskip}											
\multicolumn{5}{c}{	\underline{\makebox[0.46\textwidth][c]{TESS}}		}								\\
		2021, Aug. 21	&$	59\,447	$&	42	&	120	&	20 d	\\
		2022, Sep. 2	&$	59\,845	$&	56	&	120	&	27 d	\\
		2023, Sep. 20	&$	60\,208	$&	70	&	120	&	24 d	\\
        2024, Sep. 5    &$  60\,559 $&  83  &   120 &   24 d    \\
\hline\noalign{\smallskip}																	
    \end{tabular}
    \tablefoot{Dates and Julian dates refer to the time of first observation of a listed set.}
    \label{T:LOG:PHOT}
\end{table}

Recently, new photometric and spectroscopic data of EI~Psc became available. We used them in our analyses, and below we give a short description. 

\subsection{Spectroscopic observations}

There is a low-resolution spectrum of the EI~Psc in the LAMOST database 
\citep{2020PASJ...72...76H} obtained in 2013. It shows a blue continuum,
strong double-peaked Balmer emission lines, and \ion{He}{I} emission 
lines that are accompanied by a lot of metal absorption from the 
secondary. In 2021, the object was observed with the VLT/X-Shooter 
spectrograph~\footnote{ESO programme 0105.D-0014(A), PI: Pala, A. F.}. 
Three consecutive observations were obtained using all three arms of the
X-Shooter simultaneously (see  Table\ref{T:LOG:SP} for details). 
The observational setup provided spectra with S/N ratios in a continuum 
ranging from $\sim 40$ to $\sim 50$ in the centre of their spectral 
range, depending on the configuration.

New time-resolved spectroscopy was obtained with the 10.4~m telescope of 
the Observatorio Roque de los Muchachos (Gran Telescopio Canarias, GTC)
using the OSIRIS
instrument~\footnote{\url{http://www.gtc.iac.es/instruments/osiris/}} in 
the long-slit mode on the nights of 2023 September 4, 5, and 7 (programme:
GTC2-23BMEX~\footnote{PI: Zharikov, S.}). The log of observations of 
EI~Psc is given in Table~\ref{T:LOG:SP}. On the first night, we used 
the grism R2500I (7330 -- 10000 \AA),  on the second R2500R 
(5575 -- 7685 \AA), and on the third R2500V (4500 -- 6000 \AA),
which gave an $R \sim$ 2500 resolution. In each observational run, 
we obtained 29 spectra with an individual exposure time of 181~sec, 
allowing us to cover about 120~min, just under the length of two 
orbital periods. The standard~{\sc iraf} procedure was used to reduce 
the data. 

\begin{figure*}[ht]
    \centering
    \includegraphics[width=1.03\textwidth,clip=]{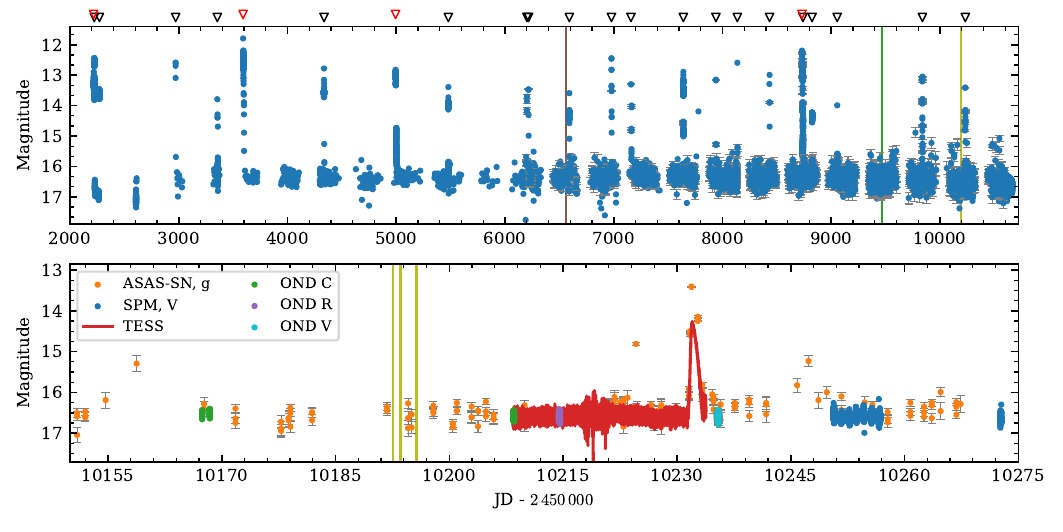}
    \caption{\textit{Top:} Long-term light curve of EI~Psc composed of 
    ASAS-SN, ZTF, Gaia, AAVSO, TESS, SPM, \ond\ observations, and 
    observations published by \cite{2002PASP..114..630S}. The vertical 
    lines represent the date of spectral observations
    of LAMOST (brown), ESO (green), and GTC (yellow). The black and red 
    triangles mark the observed outbursts and superoutbursts, 
    respectively. \textit{Bottom:} Light curve showing the scheduling of 
    different observations focusing on the time of GTC observations 
    (yellow vertical lines) and SPM observations. The photometric data in 
    both panels were shifted in such a way that the quiescence level would 
    be the same as for the SPM data.  }
    \label{F:LC:LONG}
\end{figure*}

\begin{figure}
   \centering
    \includegraphics[width=0.5\textwidth, bb = 0 0 250 192, clip=]{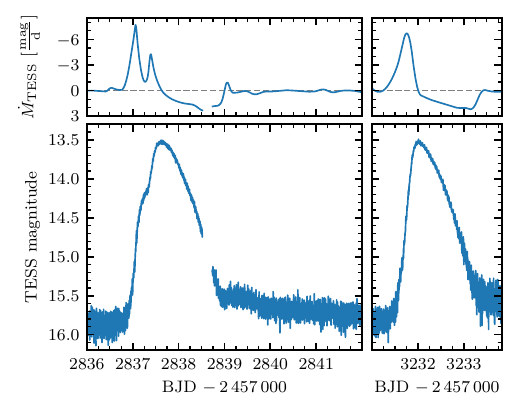}
   \caption{Outbursts observed by \T\ mission on September 3, 2022 and 
   October 2, 2023. The brightness change rate is indicated in the upper 
   panels.  }
   \label{F:TESS_OBS}
\end{figure}

\begin{figure}
   \centering
   \includegraphics[width=0.5\textwidth, bb = 10 10 250 177, clip=]{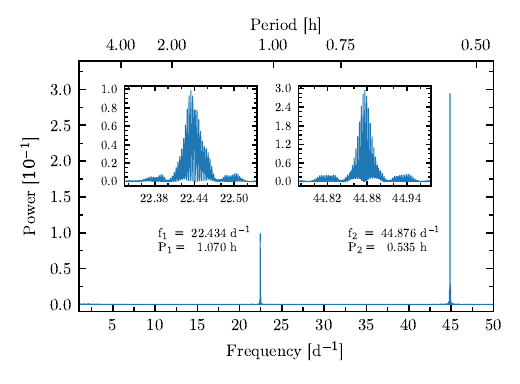}
   \caption{Periodogram of the \T\, SPM, and \ond\ data focusing on the 
   orbital period. Only the data obtained  in quiescence were used.}
   \label{F:TESS_PERIOD}
\end{figure}

\subsection{Transiting Exoplanet Survey Satellite}
EI~Psc was observed by the {\it Transiting Exoplanet Survey Satellite} 
(\T), \cite{2015JATIS...1a4003R} in four sectors. The details of the 
observations are summarised in Table~\ref{T:LOG:PHOT}.
We used the MAST~\footnote{MAST: Barbara A. Mikulski Archive for Space 
Telescopes, \url{https://mast.stsci.edu/portal/Mashup/Clients/Mast/Portal.html}} 
database to download these photometric time series.  

\begin{figure*}[t]
    \centering   
    \includegraphics[width=0.99\textwidth]{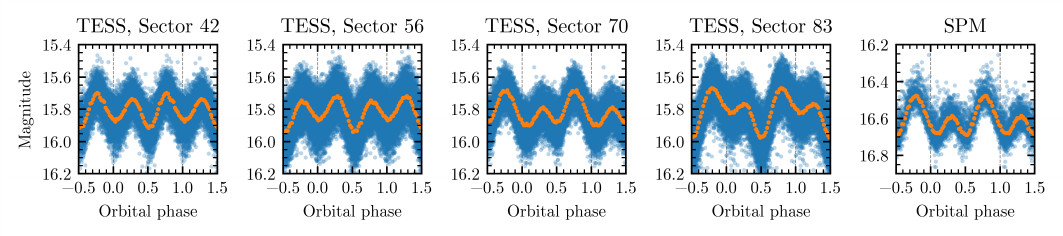}
    \caption{\T\ and SPM phase-folded light curves in quiescence. 
    Only \T\ measurements with an error smaller than 
    $5 \, \mathrm{e}^{-}\, \mathrm{s}^{-1}$. The orange points show the 
    binned light curve with the width of the bins set to $0.02$ of the 
    orbital phase.}
    \label{F:S:TESS_ALL}
\end{figure*}

\begin{figure}[htb]
    \centering   
    \includegraphics{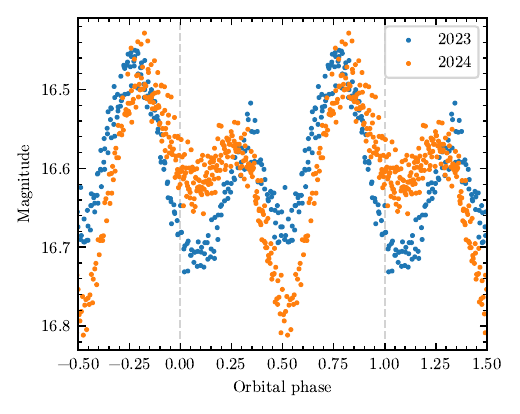}
    \caption{ Phase-folded light curves obtained at the \ond\ Observatory 
    in 2023 and 2024.}
    \label{F:S:OND_23_24}
\end{figure}

\begin{figure*}[t]
    \centering   
    \includegraphics{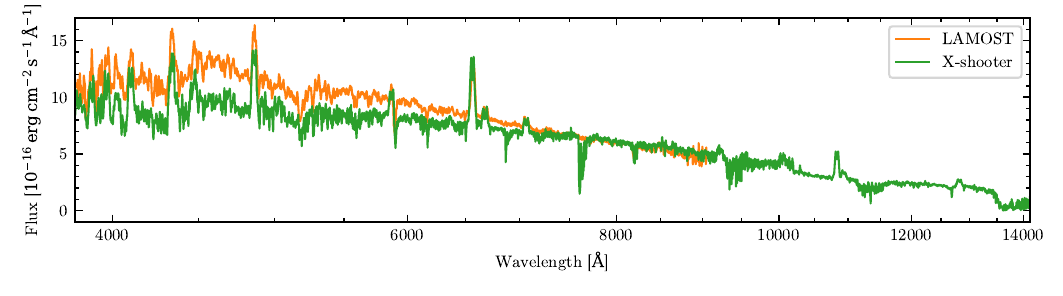}
    \caption{LAMOST and X-shooter flux calibrated spectra of EI~Psc. }
    \label{F:S:spectrum}
\end{figure*}

\subsection{Ground-based photometry}

Our new CCD observations and time series photometry were obtained at 
two observatories: Observatorio Astronomico Nacional San Pedro 
M\'{a}rtir (M\'{e}xico, hereafter OAN SPM) and \ond\ Observatory
(Czech Republic, hereafter \ond). A log of the observations is
given in Table~\ref{T:LOG:PHOT} and a short description is given 
in the following subsections.

\subsubsection{\sc OAN SPM}

Time-resolved photometry of \object{EI~Psc} was obtained in the V-band using 
the direct CCD image mode at the 0.84-m 
telescope~\footnote{\url{https://www.astrossp.unam.mx}} 
in November 2023. The images were bias-corrected and flat-fielded 
before aperture photometry was carried out. The photometric data of 
OAN SPM were calibrated using Landolt standard stars. Photometric 
errors were calculated from the dispersion of the magnitude of 
comparison stars in the object field. They range from 0.01~mag to
0.03~mag. 

\subsubsection{\ond}

Time-resolved CCD photometry of \obj\ has also been obtained at the 
\ond\ observatory, Czech Republic, in 2023 and 2024. Observations 
were carried out using the Mayer 0.65-m  ($f/3.6$) reflecting telescope
with the MI G2-3200 CCD camera mostly without a 
photometric filter (clear). The maximum quantum efficiency ($QE$) 
of the CCD chip is at $\sim$6200~\AA\ and  FWHT$_{QE}$ is about 2000~\AA. 
{\sc Aphot}\footnote{Developed at the \ond\ Observatory by M. Velen 
and P. Pravec.}, a synthetic aperture photometry and astrometry software, 
was used for data reduction. Differential photometry was performed using 
comparison stars in the object's field.

\subsubsection{Additional ground-based observations}

To study the long-term activity of EI~Psc, we use data obtained by 
the All-Sky Automated Survey for 
Supernovae\footnote{\url{https://asas-sn.osu.edu/}} \citep[ASAS-SN;
\textit{V}~and \textit{g} filters; ][]{2014ApJ...788...48S, 
2017PASP..129j4502K}, the Zwicky Transient Facility (ZTF) 
survey\footnote{\url{https://www.ztf.caltech.edu/ztf-public-releases.html} 
} \citep[\textit{i}, \textit{r},~and \textit{g} filters; ][]
{2019PASP..131a8003M}, American Association of Variable Star 
Observers\footnote{\url{https://www.aavso.org}} \citep[AAVSO; \textit{V} 
filter;][]{AAVSO:ONLINE}, and observations published by 
\cite{2002PASP..114..630S}.

\begin{figure*}
    \centering
    \includegraphics{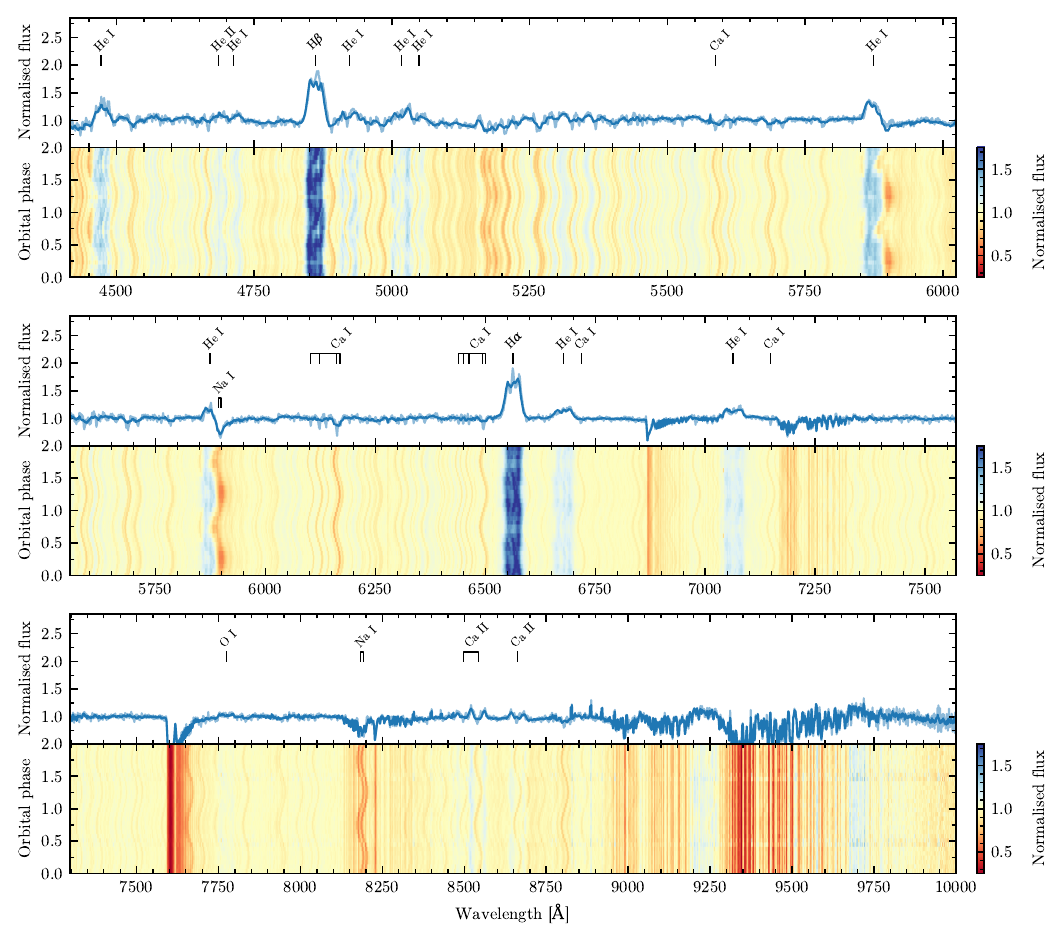}
    \caption{Spectra of EI~Psc obtained at GTC with grism R2500V 
    (top panels), R2500R (middle panels), and R2500I (bottom panels). 
    The upper panels show averaged spectra in blue and a single spectrum 
    in light blue, single spectra were obtained during the orbital phase 
    $\varphi \doteq 0.0$. The lower panels show trailed spectra.}
    \label{F:S:GTC}
\end{figure*}

\section{Light curve analyses}
\label{s:lightCurve}

\begin{table}
\caption{Registered outbursts of EI~Psc during 2001--2023.}
\label{T:OBS}
\centering
\begin{tabular}{l c r }
\hline \noalign{\smallskip}
		JD $- 2\,400\,000$	&	$T_d$ [day]	&	Type	\\
\hline\hline \noalign{\smallskip}
$	52\,217.3\tablefootmark{a}	$ &	$T_d >5 $	&	superoutburst	\\
$	52\,223.5	$ &	$T_d>1.3$	&	normal outburst	\\
$	52\,270.3	$ &		&		\\
$	52\,971.3	$ &		&		\\
$	53\,354.3	$ &	$T_d<5$	&		\\
$	53\,591.4	$ &	$T_d>9$	&	superoutburst	\\
$	54\,335.7	$ &	$T_d>2$	&	normal outburst	\\
$	54\,993.6	$ &	$T_d>2$	&	superoutburst	\\
$	55\,478.6	$ &		&		\\
$	56\,200.9	$ &	$1< T_d <2.3$	&	normal outburst	\\
$	56\,215.8	$ &	$2< T_d <4$	&	normal outburst	\\
$	56\,591.4	$ &	$1 < T_d < 4$	&	normal outburst	\\
$	56\,978.9	$ &	$3 < T_d< 7$	&		\\
$	57\,158.2	$ &	$ T_d > 3$	&		\\
$	57\,639.3	$ &	$ T_d \geq 2 $	&	normal outburst	\\
$	57\,939.0	$ &	$2 < T_d < 6$	&	normal outburst	\\
$	58\,135.8	$ &	$ T_d <4$	&	normal outburst	\\
$	58\,432.3	$ &	$3$	&	normal outburst	\\
$	58\,730.5	$ &	$10$	&	superoutburst	\\
$	58\,739.6	$ &		&		\\
$	58\,823.5	$ &		&		\\
$	59\,054.4	$ &	$1 < T_d \leq 2$	&	normal outburst	\\
$	59\,837.6	$ &	$2$	&	normal outburst	\\
$	60\,232.0	$ &	$2$	&	normal outburst	\\

\noalign{\smallskip}  \hline
\end{tabular} 
\tablefoot{$T_d$ stands for the duration of outbursts and superoutbursts.}
\end{table}

The long-term light curve of \obj collected by different photometric
sky surveys is shown in Fig.~\ref{F:LC:LONG}. The object exhibits normal 
dwarf nova outbursts and superoutbursts, which is typical for SU UMa dwarf 
novae \citep{2016MNRAS.460.2526O} as well as for AM CVn stars 
\citep{2021MNRAS.508.3275P}. The list of observed outbursts is presented 
in Table~\ref{T:OBS}.
Normal outbursts last $\sim$2 days and their amplitude is about 3 mag. 
The duration of superoutbursts is about $\sim$10 days and their 
amplitude is about 4~mag. They go through a plateau phase after 
reaching the maximal brightness. 

The light curve shown in Fig.~\ref{F:LC:LONG} covers about 8400 days 
during which 20 normal outbursts and four superoutbursts were observed. 
However, the recurrence times of outbursts cannot be reliably determined 
because of the spare coverage of the light curve and the short duration 
of outbursts. The spare coverage could also cause partially observed 
superoutbursts to be misclassified as normal outbursts. Two normal 
outbursts were observed by \T\ mission in their entirety (except for a 
small gap), the corresponding light curves are presented in 
Fig.~\ref{F:TESS_OBS}. These outbursts show an asymmetric shape with a 
fast increase in brightness ($\sim 2 \; \mathrm{mag \, day}^{-1}$) which 
is followed by a slower decrease ($\sim 1.6\; \mathrm{mag \, day}^{-1}$). 
In the upper panels of Fig.~\ref{F:TESS_OBS}, we also show the evolution 
of the rate of the brightness change during the outbursts. 
Although the shapes of both outburst profiles look similar, the first 
shows a deviation in the brightness increasing rate that occurred $0.5$ 
days after the onset of the outburst when the system reached about $2/3$ 
of the maximal brightness. 

At the timescale of the orbital period, the object shows a double-humped
light curve variation with an average amplitude of $\approx0.1$~mag in the 
quiescence. The variation is also present during outbursts, but it is less 
pronounced. We have calculated the Lomb-Scargle periodogram 
\citep{1976Ap&SS..39..447L, 1982ApJ...263..835S, 2018ApJS..236...16V} of  
\T , SPM, and \ond\ data (see Fig.~\ref{F:TESS_PERIOD}) and found two 
strong peaks ($f_1, f_2$) corresponding to the orbital period $P_{\rm orb} 
= 1.070\,\mathrm{h}$ and its half ($P_{2} = 0.535\,\mathrm{h}$). 
The peak corresponding to half of the orbital period is stronger because 
of the double-humped nature of the light curve. By fitting the local 
maxima of the peaks with a Gaussian profile, we derived two estimates of 
the orbital period: \begin{equation} 
   \begin{aligned}
    P_{\mathrm{orb, f1}} &= 0.044571(32)~\mathrm{d}, \\
    P_{\mathrm{orb, f2}} &= 0.044567(13)~\mathrm{d}.
    \end{aligned}
    \end{equation} 
Both estimates agree within the error and also with the value of the 
orbital period reported by \citet{2002ApJ...567L..49T}. In further 
analysis, we adopted the value $P_{\mathrm{orb, f2}}$, as it could be 
determined with higher accuracy. 

The optical light curves, folded on the orbital period, are shown in 
Fig.~\ref{F:S:TESS_ALL} and Fig.~\ref{F:S:OND_23_24}. The minima of the 
light curves are located in the spectroscopic phase $\phi$ (see below) of 
0.0 and 0.5 and the maxima at phases 0.25 and 0.75. The light curve was 
symmetric in the optical and infrared data reported by 
\citet{2002ApJ...567L..49T, 2009AJ....137.4061H}, and in the first two 
sets of \T\ observations. 
However, the light curve exhibits an asymmetric shape in the last two 
sectors of \T\, SPM observations and \ond\ observations. The amplitude 
of the peak at $\phi = 0.75$ is twice as large as the amplitude of the 
peak at $\phi = 0.25$ and the light curves from 2024 also show minima of
different depths.

\section{Spectroscopy}
\label{S:Spectra}

Fig.~\ref{F:S:spectrum} shows the flux-calibrated X-Shooter spectrum 
(green line) of EI~Psc together with the LAMOST spectrum (orange line) 
shifted to the X-Shooter spectrum in its red limit. The LAMOST spectrum 
shows a bluer continuum compared to the X-Shooter, which can be explained 
by the fact that it was obtained closely before an outburst. Both spectra 
clearly show absorption components originating from the secondary. A 
visible presence of WD contribution is absent in both spectra.

Fig.~\ref{F:S:GTC} shows normalised trailed GTC spectra of EI~Psc 
together with normalised phase-averaged spectra.
Various emission and absorption lines are marked. The absorption lines 
clearly show a sinus-like variation in their radial velocities. The 
emission lines are double- or multi-peaked and their trailed spectra 
show an S-wave, which varies close to the opposite of absorption's 
radial velocities. 

\begin{figure}
    \centering
    \includegraphics{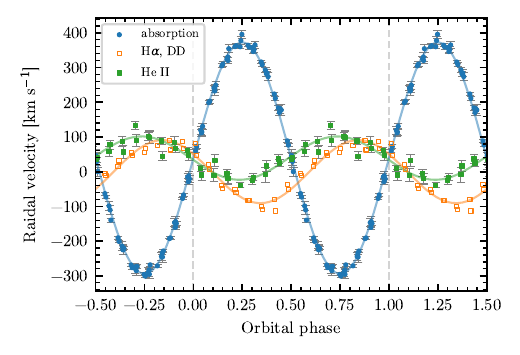}
    \caption{
    Radial velocities of absorption lines and  H$\alpha$ and 
    \ion{He}{II} emission lines. The solid lines show the best sinusoidal
    fits of each set, parameters of the best fits are given in 
    Table~\ref{T:VC}.}
    \label{F:P:FLC}
\end{figure}

We measured the radial velocities (RVs) of the wings of the H$\alpha$ 
line using the diagnostic diagram method (DD) described by 
\cite{1986ApJ...308..765S}. The resulting radial velocity curve is shown 
in Fig.~\ref{F:P:FLC} and the best-fit parameters are presented in 
Table~\ref{T:VC}. This method produces a radial velocity curve that 
corresponds to the inner part of the accretion disc and can therefore be 
used to track the radial velocities of the primary. However, the 
reliability of this method depends on the symmetric brightness 
distribution in the inner parts of the disc and can produce unrealistic 
measurements when there is an asymmetry \citep[see][chap. 2.7.6]
{1995cvs..book.....W}.

As absorption features of the secondary are prominent in all newly 
obtained spectra, we used them to measure its radial velocities using the 
cross-correlation method (CC). For this, we used the FXCOR task in {\sc 
iraf}. We used the whole range for each set of spectra, excluding the 
parts containing spectral features corresponding to the primary, accretion 
disc, and telluric lines. The first spectrum of each set was used as the 
reference spectrum; therefore, the measured velocities are only relative 
values. For further analysis, we shifted the measured radial velocities so 
that they can be described by a sine curve with zero vertical shift; the 
resulting radial velocity curve is shown in Fig.~\ref{F:P:FLC}, parameters 
of the best fit are presented in Table~\ref{T:VC}. We confirm the presence 
of a small phase displacement between the zero phase of absorptions and 
the Balmer lines reported in the first by \citet{2002ApJ...567L..49T}.

\begin{figure}
    \centering
    \includegraphics{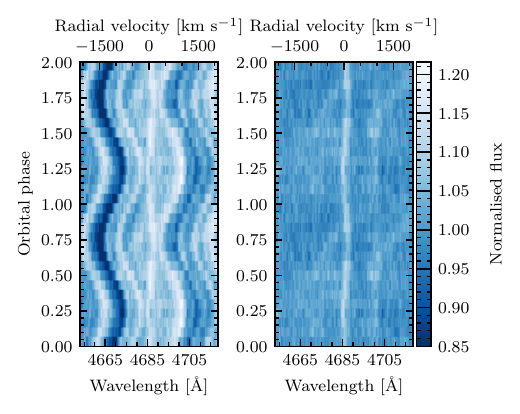}
    \caption{Trailed spectra zoomed in on \ion{He}{ii} 
    4686~\AA. The left plot shows raw data and the right one 
    presents the spectra after removing the contribution of the secondary.
    } 
    \label{F:TS:HeII}
\end{figure}

We also determined the radial velocity of \ion{He}{II} 4686 \AA\ line. 
As shown in Fig.~\ref{F:TS:HeII},~left it is weak and surrounded by strong 
absorption lines from the secondary. 
To exclude the contribution from the secondary, we computed its average 
spectrum by transforming the spectra into its rest frame and masking out 
the \ion{He}{II} emission. Then, we subtracted the average spectrum from 
the observed one and measured \ion{He}{II} RV by fitting its profile with 
a Gaussian function. The trailed spectra of \ion{He}{ii} 4686 \AA\ line 
after absorption subtraction are presented in Fig.~\ref{F:TS:HeII},~right.
The results of the RV measurements are presented in Fig.~\ref{F:P:FLC}  
and parameters of the best fit are given in Table~\ref{T:VC}.
We found that \ion{He}{II} moves completely in the anti-phase of the 
absorption lines, from which we conclude that the \ion{He}{II} 4686 \AA\ 
line originates at the primary. The RV curve of H$\alpha$ measured by the 
DD method is not completely in anti-phase with the RV curve of the 
secondary, which could be attributed to the non-uniform distribution of 
radiation from the disc, and this could also affect the amplitude of the 
RV curve. Therefore, we conclude that the RV curve of \ion{He}{II} is a 
better approximation of the RV curve of WD.

\begin{figure}[t]
    \centering   
    \includegraphics{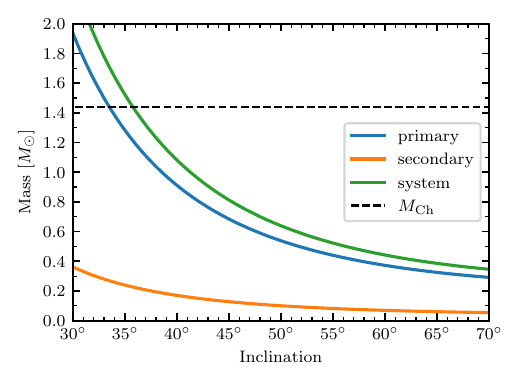}
    \caption{Relationships for system components masses vs. of the
    inclination. The green line corresponds total mass of the system. 
    The primary and the secondary masses are shown by blue and red lines,
    respectively. The dashed line denotes the Chandrasekhar limit 
    M$_{\rm Ch} = 1.44$ \ms.}
    \label{F:MF}
\end{figure}

The mass ratio $q=K_{\rm He II}/K_2$ = 0.187(14) is practically the same 
as the estimations of \citet{2002ApJ...567L..49T}. 
Therefore, if we adopt $K_{\rm WD} \equiv K_{\rm He II}$ the mass 
functions (see Fig.~\ref{F:MF}) are
\begin{equation}
     M_{\rm WD} \sin^3 i = \frac{P_{\rm orb} K_2 (K_{\rm WD} + K_2)^2}{2\pi G} = 0.24(1)~M_\odot,
\end{equation}

\begin{equation}
M_{2} \sin^3 i  =  \frac{P_{\rm orb} K_{\rm WD} (K_{\rm WD} + K_2)^2}{2\pi G}  =  0.0045(4)~M_\odot,
\end{equation}
and the system separation (a) is 
\begin{equation}
a \sin i = \frac{P_{\rm orb} (K_{\rm WD} + K_2)}{2\pi} = 0.349(5)~R_\odot,
\end{equation}
where $M_{\rm WD}$ and $M_{2}$ are masses of WD and the secondary, $i$ is 
the system inclination, and $G$ is the gravitational constant.

 The ephemeris of the system based on the radial velocity curve of the
 secondary was derived as 
 \begin{equation}
\mathrm{}{T}_0 = \mathrm{BJD}\ 2\,460\,193.556926(25) + 0.044567(13) \cdot E,
\end{equation} where $\mathrm{}{T}_0$ is the zeroth phase and the orbital
period was determined from the \T, OAN SPM, and \ond\ observations.
We searched for a variation of equivalent widths (EW) of absorption with 
the orbital period and found that most of the lines show small dispersion 
($\lesssim$ 20\%) of EW, which weakly depends on the system orbital phase. 
This can be interpreted as a homogeneous distribution of the surface 
temperature of the secondary.

\begin{figure}[t]
    \centering   
    \includegraphics{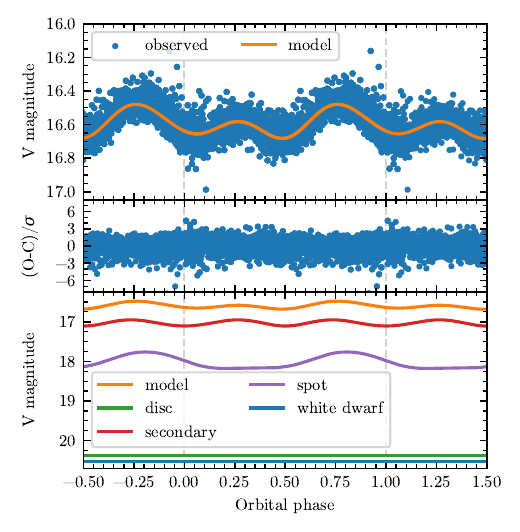}
    \caption{Top panel: V-band light curve (blue dots) folded to the
    orbital period and its simulation result. Middle panel: Observed minus 
    calculated (O-C) plot in terms of data uncertainty. Bottom panel: 
    Contribution of each model component.}
    \label{F:LCmod}
\end{figure}

\begin{figure}[t]
    \centering   
    \includegraphics[width=0.5\textwidth, bb = 20 10 1490 1300, clip= ]{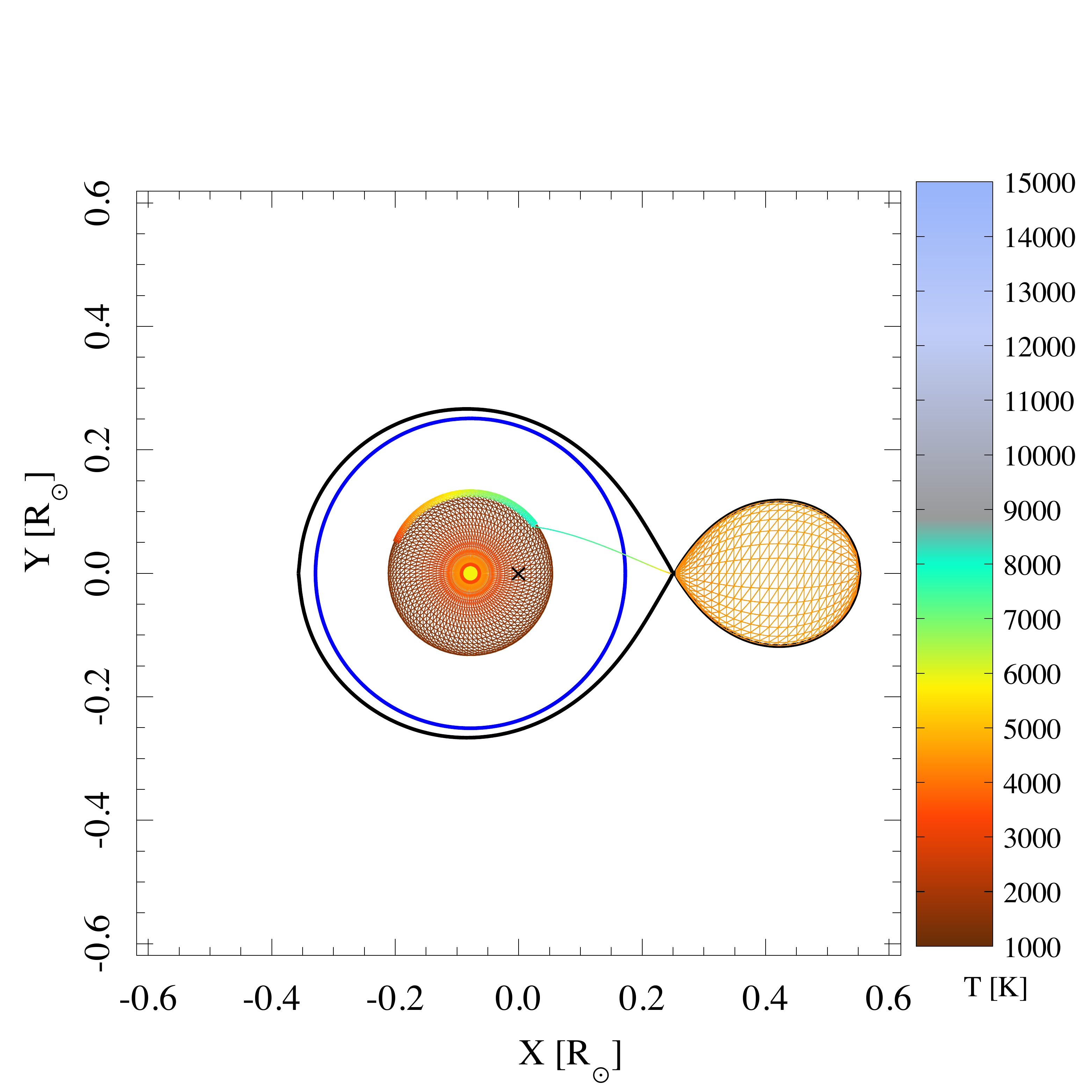}  
    \caption{Geometry of EI~Psc. The colour bar corresponds to the 
    effective temperature of the emitting regions. The blue circle shows 
    the disc tidal truncated radius. The cross marks the mass centre of 
    the system.}
    \label{F:Zero}
\end{figure}

\begin{table}
\caption{Radial velocity solution}
\label{T:VC}
\centering
\begin{tabular}{llll}
\hline \noalign{\smallskip}
Method      &  $\upsilon_{\rm cor}$ [$\mathrm{km \, s}^{-1}$]  & $K$  [$\mathrm{km \, s}^{-1}$] & $\varphi_0$ \\
\noalign{\smallskip}  \hline \hline
\noalign{\smallskip}  
DD   & $0.1(2.9)$ & $90.9(4.2)$  & $0.093(7)$ \\
CC   &           & $-333.9(1.1)$ & $0.0006(6)$ \\ 
\ion{He}{II} &  $39.1(3.2)$   &  $62.6(4.6)$    &  $-0.017(11)$  \\
\noalign{\smallskip}  \hline
\end{tabular} 
\begin{tablenotes}
 \item[] DD stands for "diagnostic diagram"; CC stands for "cross-
 correlation of the absorption lines";
 $\upsilon_{\rm cor}$ is $\gamma$-velocity of the system, $K$ is the 
 amplitude of a RV curve, and $\varphi_0$ is its zeroth phase
\end{tablenotes}
\end{table}

\begin{table}
\label{T:SysPar}
\centering
\caption{ System parameters from the light-curve modelling of 
EI~Psc in quiescence.
}
\label{tab:BestPar}
\begin{tabular}{llll}
\hline\noalign{\smallskip}
{\bf Fixed parameters:}       &            &          \\  
\hline\hline\noalign{\smallskip}
\multicolumn{2}{l}{$P_{\mathrm{orb}}$}  & \multicolumn{2}{c} {3850.8 s}   \\ 
\multicolumn{2}{l}{$E(B-V)$ }           & \multicolumn{2}{c} {$0.00^{+0.01}_{-0.00}$}    \\
\multicolumn{2}{l}{Distance}            & \multicolumn{2}{c} {148.5(1) pc}  \\   
\multicolumn{2}{l}{$q = M_2/M_{\rm WD} $}  & \multicolumn{2}{c}{ 0.187}  \\
\multicolumn{2}{l}{$T_{2}$ }  & \multicolumn{2}{c}{$ 4440(100)$  K}\\
\hline\noalign{\smallskip}
\multicolumn{4}{l}{{\bf Variable and their best values:}} \\  \hline\noalign{\smallskip}
\multicolumn{2}{l}{ $i$ }               & \multicolumn{2}{c}{44\fdg5(6.5) } \\
\multicolumn{2}{l}{$M_{\mathrm{WD}} $ } &\multicolumn{2}{c}{0.70(4) M$_{\sun}$ }  \\
\multicolumn{2}{l}{${T}_{\mathrm{WD}}$} & \multicolumn{2}{c}{6130$^{+1500}_{-4000}$ K}\\ 
\multicolumn{2}{l}{$\dot{M}$$\times10^{-13}$M$_{\sun}$ year$^{-1}$} & \multicolumn{2}{c}{3.9(6)}\\

\hline\noalign{\smallskip}
\multicolumn{4}{l}{{\bf Parameters$^*$ of the disc:}} \\ 
\hline\noalign{\smallskip}
\multicolumn{2}{l}{ $R_{\mathrm{d, in}}$  } & \multicolumn{2}{c}{$\equiv R_{\rm WD}$ } \\
\multicolumn{2}{l}{ $R_{\mathrm{d, out}}$ } & \multicolumn{2}{c}{ 0.134(23) R$_{\sun}$ } \\
\multicolumn{2}{l}{ $h_{\mathrm{d, out}}$ } & \multicolumn{2}{c}{ 0.003(3)  R$_{\sun}$ } \\  

\noalign{\smallskip}\hline\noalign{\smallskip}
\multicolumn{2}{l}{{\bf The hot spot/line:}}&  \multicolumn{2}{c}{ }  \\ \hline\noalign{\smallskip}
\multicolumn{2}{l}{length spot ($\varphi_{\mathrm{min}}+ \varphi_{\mathrm{max}}$) }  &\multicolumn{2}{c}{123(17) }  \\
\multicolumn{2}{l}{width spot (\%) }  &\multicolumn{2}{c}{0.064(14) } \\
\multicolumn{2}{l}{Temp. excess spot ($T_{\mathrm{s, max}}/T_{\mathrm{d,out}}$)  }  &\multicolumn{2}{c}{4.65(15) } \\
\hline\noalign{\smallskip}
\multicolumn{4}{l}{{\bf Calculated parameters:}} \\ 
\hline\noalign{\smallskip}
\multicolumn{2}{l}{$a $}  &\multicolumn{2}{c}{0.50  R$_{\sun}$}  \\
\multicolumn{2}{l}{$M_{2}$}  & \multicolumn{2}{c}{0.130 M$_{\sun}$} \\
\multicolumn{2}{l}{$R_{2}$}  & \multicolumn{2}{c}{0.123 R$_{\sun}$}\\
\multicolumn{2}{l}{$R_{\mathrm{WD}}$ } & \multicolumn{2}{c}{0.011 R$_{\sun}$}\\
\hline

\end{tabular}
\tablefoot{Numbers in brackets are uncertainties of variables. 
Distance and $E(B-V)$ are given with 1$\sigma$ errors. The error of the 
effective temperature of the secondary corresponds to the uncertainty 
of the spectral class determination.
 }
\end{table}

\section{Quiescence light curve modelling}
\label{S:LCM}

 We used the V-band light curve to apply a modelling tool developed 
 by \citet{2013A&A...549A..77Z} and described in detail in 
 \citet{2020MNRAS.497.1475S} and \citet{2021A&A...652A..49K, 
 2023ApJ...950...47K} to redetermine the system parameters and to study 
 the structure of the accretion disc. The model assumes that the disc 
 radiates as a black body at the local effective temperature with radial 
 distribution across the disc given by the equation:
\begin{equation}
 \begin{aligned}
T_{\rm eff}(r)   &= T_0 \left\{ \left(\frac{r}{R_{\rm WD}}\right)^{-3} \, \left(1-
\left[\frac{R_{\rm WD}}{r}\right]^{1/2}\right)\right\}^{EXP},  \\
 T_0  &=  \left[ \frac{3G\,M_{\rm WD}\,\dot{M}}{8\pi\sigma\, R_{\rm WD}^3} \right]^{1/4}, 
 \end{aligned}
\label{Tempeq}
\end{equation}

\noindent
where  
$R_{\mathrm{WD}}$ 
is the radius of the non-rotating helium WD \citep{1972ApJ...175..417N}, 
$\dot{M}$ is the mass transfer rate, and
$\sigma$ is the Stefan--Boltzmann constant. 
 In the standard accretion disc model, the radial temperature gradient 
 is taken as $EXP$ = 0.25 \citep[equation 2.35]{1995cvs..book.....W},
but here we allow it to slightly deviate from this value in a manner 
similar to \citet{2010ApJ...719..271L}. The accretion disc  thickness 
is defined as
\begin{equation}
z_\mathrm{d}(r) = z_\mathrm{d}(r_{\mathrm{out}})(r/r_{\mathrm{out}})^{\gamma_{\mathrm{disc}}},
\end{equation}
where $\gamma_{\mathrm{disc}}$ is a free parameter for which we used the 
standard value of $\gamma_{\mathrm{disc}}= 9/8$ \citep[equation 2.51b]
{1995cvs..book.....W} as the initial value. We also used the complex model 
for the hot spot described in detail in \citet[see their fig.~9 and 
therein]{2021A&A...652A..49K}, where all parameter definitions can be 
found. The disc limb-darkening used in the model follows the Eddington 
approximation \citep{1980MNRAS.193..793M, 1980AcA....30..127P}.

We fit the light curve by searching for the minimum of the $\chi^2$ 
function, setting all parameters free, and using the gradient-descent 
method to obtain self-consistent results for the fit.  The best-fit 
parameters are given in Table~\ref{tab:BestPar} and the resulting light 
curves and the contribution of each component are presented in 
Fig.~\ref{F:LCmod}. The numbers in brackets for the variables in 
Table~\ref{T:SysPar} are uncertainties defined as 1$\sigma$ of the 
Gaussian function approximation used to describe the one-dimension 
$\chi^2$ function. The model predicts that in the V-band the light curve 
is formed by the combination of flux from two main components: the 
secondary and the hot spot. The contribution in the continuum of the disc 
and WD is more than ten times lower in comparison with them, indicating an 
optically thin accretion disc with an effective temperature of radiation 
only about $\sim$1500 K. The size of the outer accretion disc is only half 
of the expected tidal truncated radius of the disc. The disc luminosity of 
the system $L\approx1.1\times10^{30}$~erg~s$^{-1}$ corresponds to a low 
mass transfer accretion rate of $\dot{M} \approx 4\times 10^{-13}$ 
M$_\odot$ yr$^{-1}$.

The geometry of the system is shown in Fig.~\ref{F:Zero} for a better 
understanding of the system and disc parameters. The colour bar in the 
figure marks the effective temperature of the black-body, which 
corresponds to radiation from the system components.

\begin{figure*}
    \centering
    \includegraphics[width=0.8\textwidth]{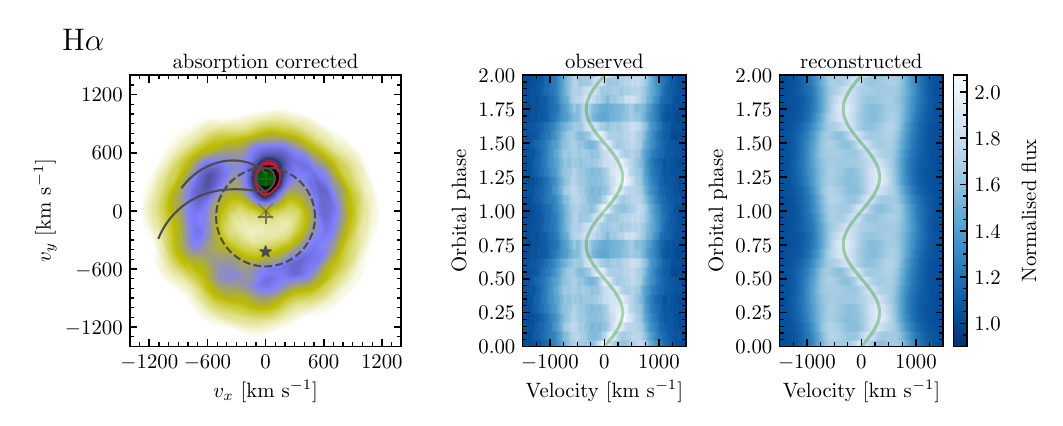}
    \includegraphics[width=0.8\textwidth]{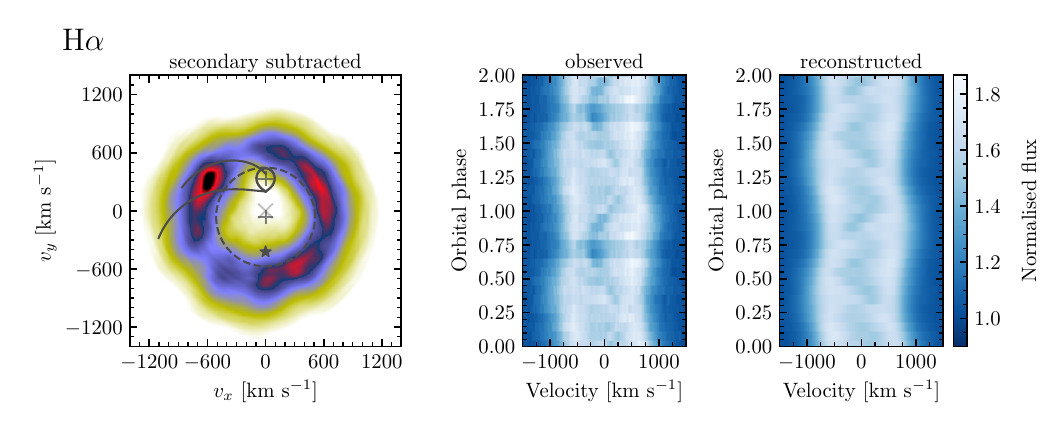}
    \includegraphics[width=0.8\textwidth]{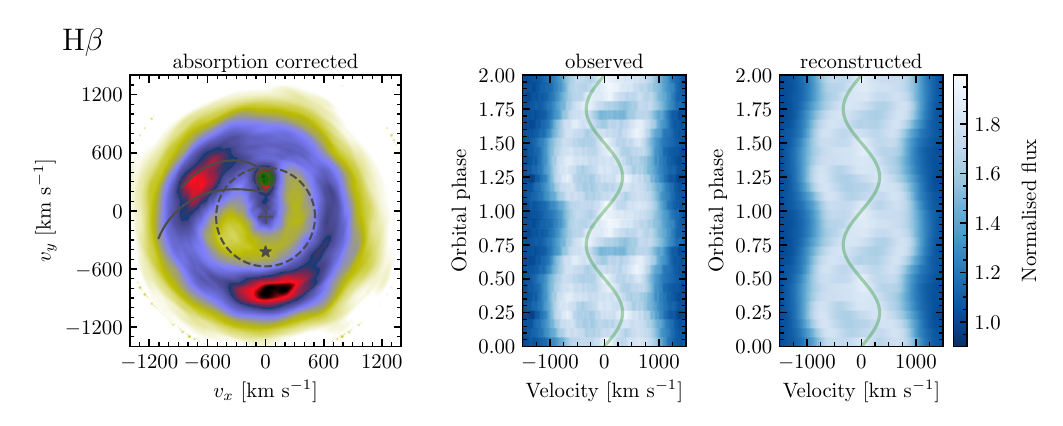}
    \caption{Absorption-corrected Doppler map of EI~Psc based on H$\alpha$ 
    (\textit{top}) and H$\beta$ (\textit{bottom}) observations obtained by GTC. 
    The Roche lobe of the secondary is outlined by solid lines, the tidal 
    limitation radius is outlined by the dashed circle. The plus signs mark the 
    positions of the stellar components, the cross marks the centre of mass, and 
    the star symbol marks the position of the $\mathrm{L}_3$ point.  The middle 
    panel shows the Doppler map based on H$\alpha$ with the peak located at the 
    position of secondary removed for better visualisation of the accretion disc 
    structure. The features related to artefacts of the absorption subtraction are 
    marked by the green circle in the maps and by the green line in the trailed 
    spectra.} 
    \label{F:DM:AC:H}
\end{figure*}

\begin{figure*}
    \centering
     \includegraphics[width=0.8\textwidth]{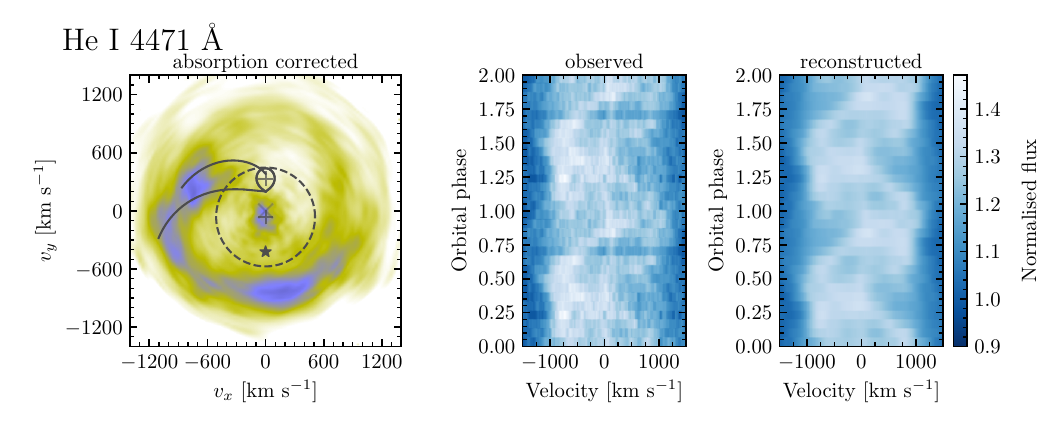}
     \includegraphics[width=0.8\textwidth]{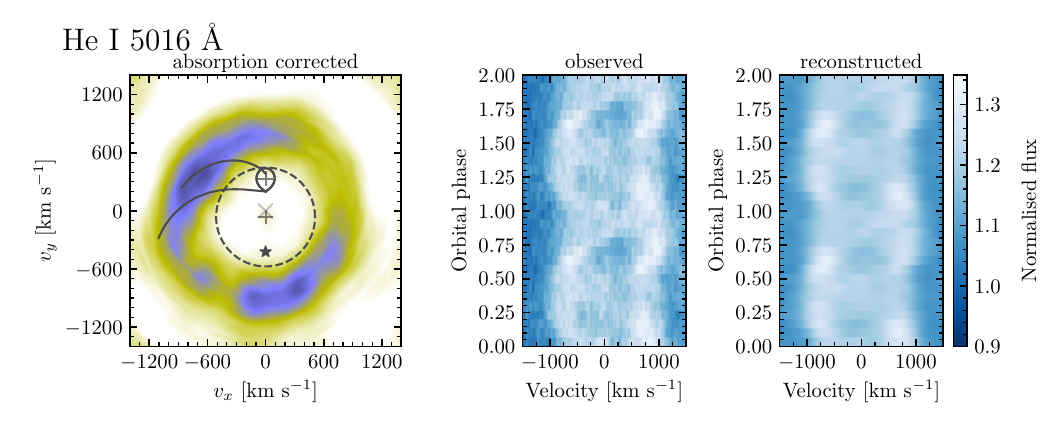}
     \includegraphics[width=0.8\textwidth]{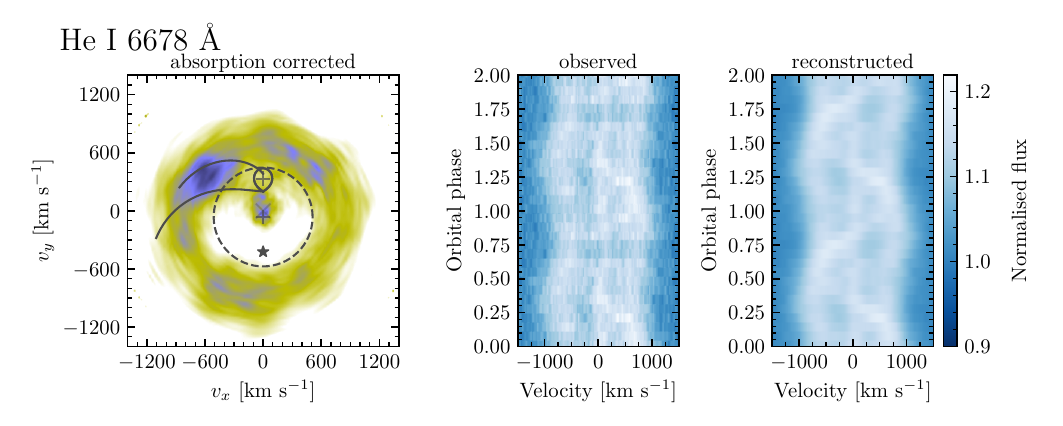}
    \caption{Absorption-corrected Doppler maps and trailed spectra  of
    \ion{He}{i}  emission lines.  See Fig.~\ref{F:DM:AC:H} for a detailed 
    description of shown elements.}
    \label{F:HeI}
\end{figure*}

\begin{figure*}
    \centering
     \includegraphics[width=0.8\textwidth]{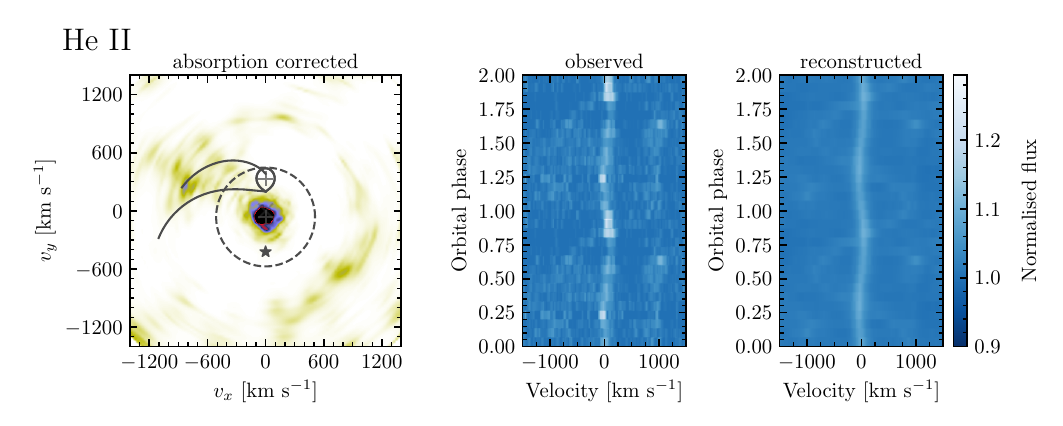}
     \includegraphics[width=0.8\textwidth]{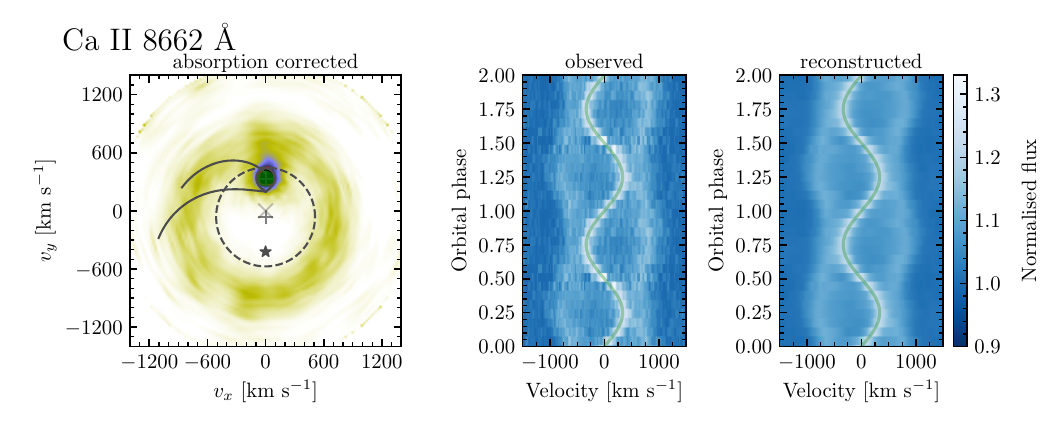}
    \caption{Absorption-corrected Doppler maps and trailed spectra  of 
    \ion{He}{ii} emission line and \ion{Ca}{ii} emission lines.  See 
    Fig.~\ref{F:DM:AC:H} for a detailed description of shown elements.}
    \label{F:HeII,CaII}
\end{figure*}

\begin{figure*}
    \centering
     \includegraphics[width=0.475\textwidth, bb = 87 65 575 479, clip=]{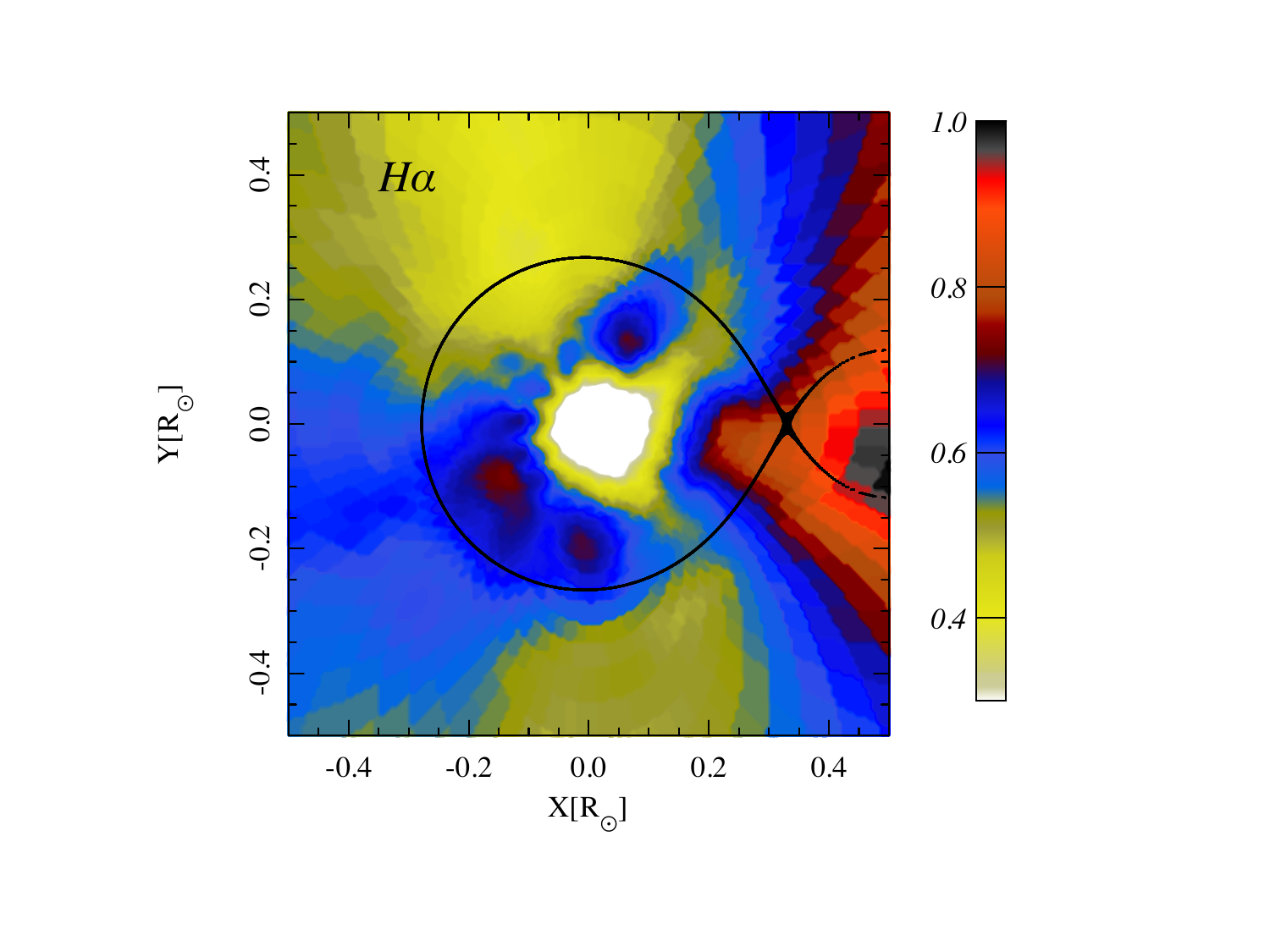}
    \includegraphics[width=0.475\textwidth, bb = 87 65 575 479, clip=]{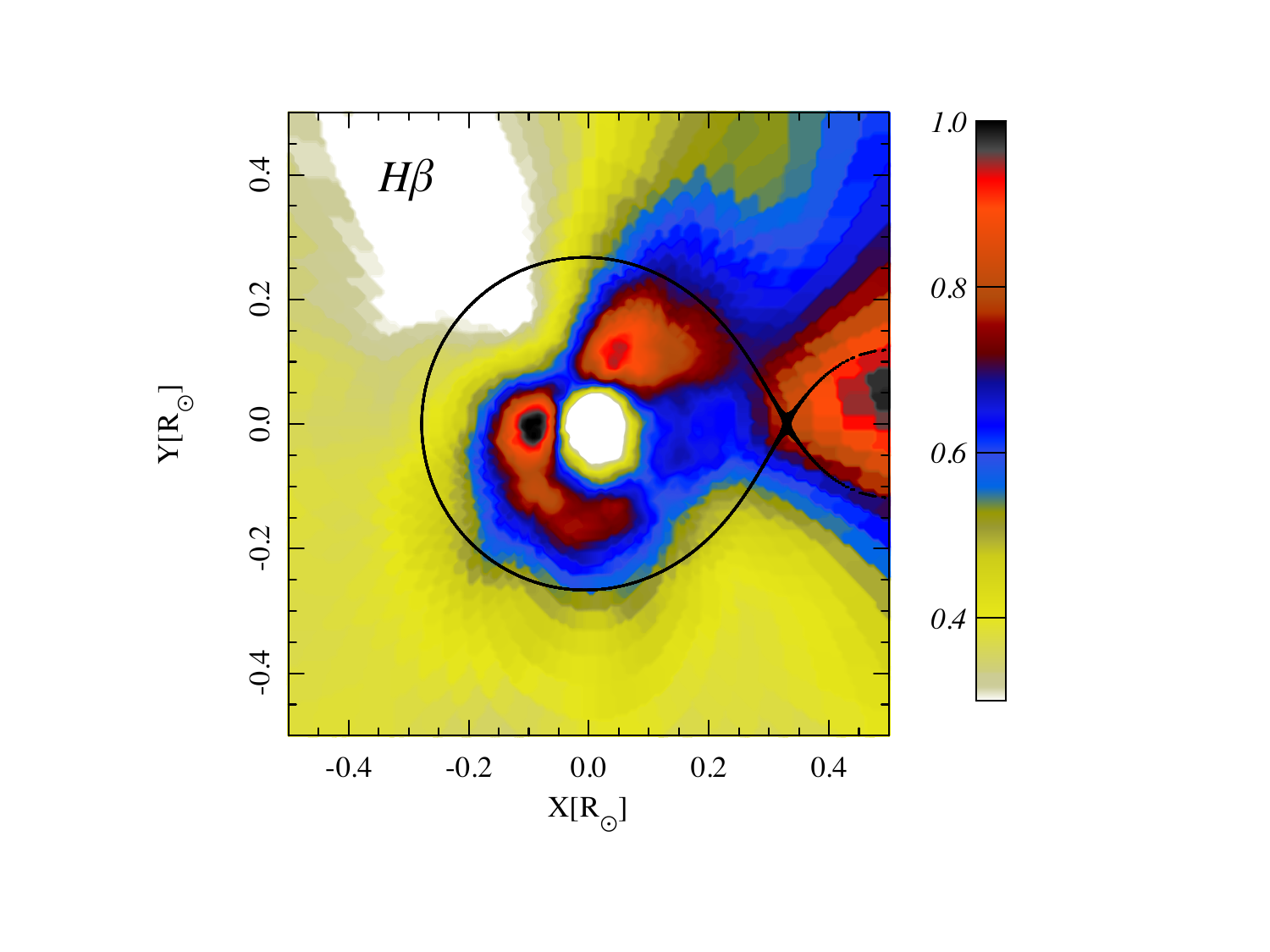}
    \caption{Accretion disc brightness distribution transformed from  H$\alpha$ 
    (left) and H$\beta$ (right) Doppler maps to the XY plane of the system. } 
    \label{F:XY:Hb}
\end{figure*}

\section{Doppler tomography}
\label{S:DT}

We used Doppler tomography (\cite{1988MNRAS.235..269M}) and GTC time-resolved 
spectra to study the structure of the accretion disc and the accretion flow in the 
system. We used the code developed by \cite{1998astro.ph..6141S} to calculate the 
Doppler maps, which we then plotted using the {\sc Python3} programming language 
and module {\sc PyDoppler} by \cite{2021ascl.soft06003H}.

As can be seen in Fig.~\ref{F:S:GTC}, spectra of the secondary exhibit numerous 
absorption lines that blend with the emission lines observed in the spectra. As 
these absorption lines can result in artifacts in the Doppler maps, we corrected 
the observed spectra for absorption before computing the Doppler 
maps\footnote{Examples of uncorrected Doppler maps are presented in 
Fig.~\ref{AtlasB}.}. We did so by subtracting an average spectrum of the K4V red 
dwarf with metallicity $\left[\mathrm{Fe/H}\right] = 0.0$ published by 
\cite{2017ApJS..230...16K}. We transformed the model spectrum from the vacuum 
wavelengths to the air wavelengths using the formula given by  \citet[see eq. 8]
{2000ApJS..130..403M} which uses the dispersion formula of 
\cite{1994Metro..31..315B}. 
To subtract the absorption lines from the observed spectra, we normalised the
model spectrum and shifted it according to a velocity correction derived from 
cross-correlation with absorption lines in the vicinity of the studied emission 
lines. The regions containing emission lines and telluric lines were not included 
in the cross-correlation analysis. Although this method proved to be effective in 
eliminating artefacts caused by neighbouring absorption lines, it also introduced 
an artificial emission at the position of the secondary in the maps based on H and 
Ca lines. This can be attributed to the difference in abundances of elements 
between the secondary in EI~Psc and an average red dwarf.
For the \ion{He}{II} line, we use  absorption-corrected spectra which we obtained 
from the procedure described in Section~\ref{S:Spectra}. 

In Figs.~\ref{F:DM:AC:H}, \ref{F:HeI}, and \ref{F:HeII,CaII}, we show selected 
absorption-corrected Doppler maps. Additional Doppler maps are shown in 
Fig.~\ref{AtlasA} and Fig.~\ref{AtlasB} show examples of Doppler maps computed 
using spectra before the absorption correction was applied. To overplot the Roche 
lobe of the secondary, the stream ballistic and Keplerian trajectories, the disc 
truncated radius, and the position of the components, we used the system 
parameters obtained from the light curve modelling (Section~\ref{S:LCM}).

Doppler maps of H$\alpha$ and H$\beta$ are presented in Fig.~\ref{F:DM:AC:H}. 
The maps show an uneven doughnut-shaped structure corresponding to the accretion 
disc. Both maps also exhibit an artifact of absorption subtraction in the form of 
a spot at the position of secondary. However, some emission in H$\alpha$ 
originating from the secondary can also be seen in the unchanged spectra (see 
Fig.~\ref{AtlasB}). The middle panel of Fig.~\ref{F:DM:AC:H} shows a Doppler map 
with the spot at the secondary position completely removed to better visualise the 
structure of the disc. The maps also show the hot spot, where the accretion stream 
impacts the disc. The hot spot is located at  co-ordinates $v_x \approx -600$ 
km~s$^{-1}$, $v_y \approx -300$ km~s$^{-1}$. This position corresponds to the 
velocity in the disc of $v_{\mathrm{spot}} \approx 700$ km~s$^{-1}$ and is in 
agreement with the velocity of the outer rim of the disc $v_{\rm disc, out} 
\sin i = 699$ km~s$^{-1}$ obtained from the light curve modelling. The H$\beta$ 
map also shows a strong emission from the part of the disc opposite the secondary 
that lies near the L$_3$ point and another emission originating at the WD or in 
its vicinity.

The \ion{He}{I} line maps (Fig.~\ref{F:HeI}) are generally similar to each other
showing similar features with varying intensity. They show the doughnut-like 
structure of the disc with increased emission coming from the hot spot and from 
the part of the disc opposite to the secondary. Some \ion{He}{I} also show low-
velocity emission originating at the position of the WD.
In contrast to \ion{He}{I}, the \ion{He}{II} map (Fig.~\ref{F:HeII,CaII}, top 
panel) shows no emission coming from the accretion disc, apart from a weak hint of 
the hot spot. The main source of \ion{He}{II} emission lies at the position of the 
WD suggesting that it originates at the WD or in its vicinity; for example, in the boundary layer between the WD surface and the accretion disc. 
The strongest feature in \ion{Ca}{II} (Fig.~\ref{F:HeII,CaII}, bottom panel) is 
the artefact of the absorption correction at the position of the secondary. Apart 
from this artefact, the map shows a doughnut-shaped structure corresponding to the 
emission from the accretion disc.

The emission originating closely to the position of the primary is presented on 
several maps. Although this feature is observed in many AM CVn stars, it is 
unusual for typical CVs. There is only one other known CV with this feature: 
the short-orbital period ($P_{orb}$ = 65~min) helium-rich dwarf nova CRTS 
J112253.3-111037 (see \cite{2012MNRAS.425.2548B} and references therein).
 
In Fig.~\ref{F:XY:Hb}, we plot the result of transforming the calculated Doppler
map into the system of spatial  co-ordinates using the assumption of the Keplerian 
velocity distribution of the emitting particles. The colour bar corresponds to the 
intensity of the Doppler map. 
The hot spot and an extended structure on the opposite side of the disc from the 
hot spot are visible. The additional radiation from the secondary (L$_1$) in the 
plots is probably caused by uncertainties in the subtraction procedure of the 
secondary contribution from observational spectra.

\begin{figure}[t]
    \centering   

    \includegraphics[width=0.4975\textwidth, bb = 0 6 740 598, clip=]{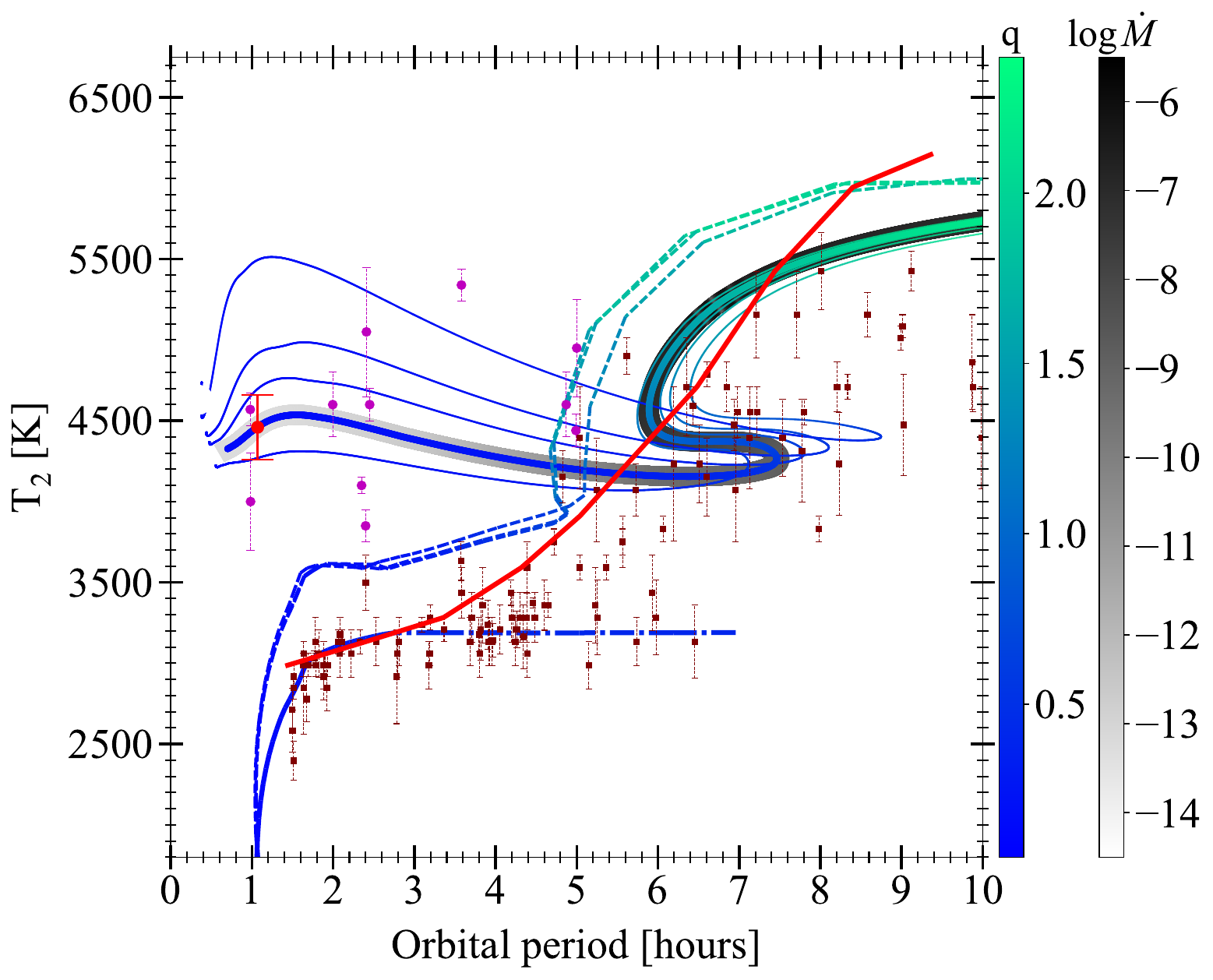} 
    \caption{ Evolution tracks of EI~Psc. The red circle marks EI~Psc position. 
    The violet circles denote known systems with warm donors for orbital periods  
    $\lesssim$~5 hours (see Table~\ref{T:WarmSecondary}). The brown squares 
    present the  effective temperatures of the donors of CVs plotted vs. their 
    orbital periods (an update of the \citet{1998A&A...339..518B} list).
    The solid red line corresponds to a relationship that assumes that the 
    secondaries follow the empirical mass-radius relation for main-sequence stars 
    based on Eq. (2.87) from \citet{1995cvs..book.....W}.
    The solid tracks show the evolutions of systems that form warm secondaries. 
    The wider track corresponds to the evolution of EI~Psc system. The dashed 
    tracks plot some of the systems with massive WDs that demonstrate close to 
    standard CV evolution leading to  CVs period minimum and forming of period 
    bouncers. The dash-point track shows an example of a conservative mass 
    transfer evolution of a cataclysmic variable with 0.7~M$_\odot$ WD mass and 
    the low mass of the secondary of 0.5~M$_\odot$. The mass transfer rate 
    $\dot{M}$ is given in M$_\sun$\ year$^{-1}$ unit.
    See details in the text. }
    \label{F:Evmodel}  
\end{figure}

\begin{figure}[t]
    \centering   
    
    \includegraphics[width=0.5\textwidth, bb = 0 0 800 650, clip=]{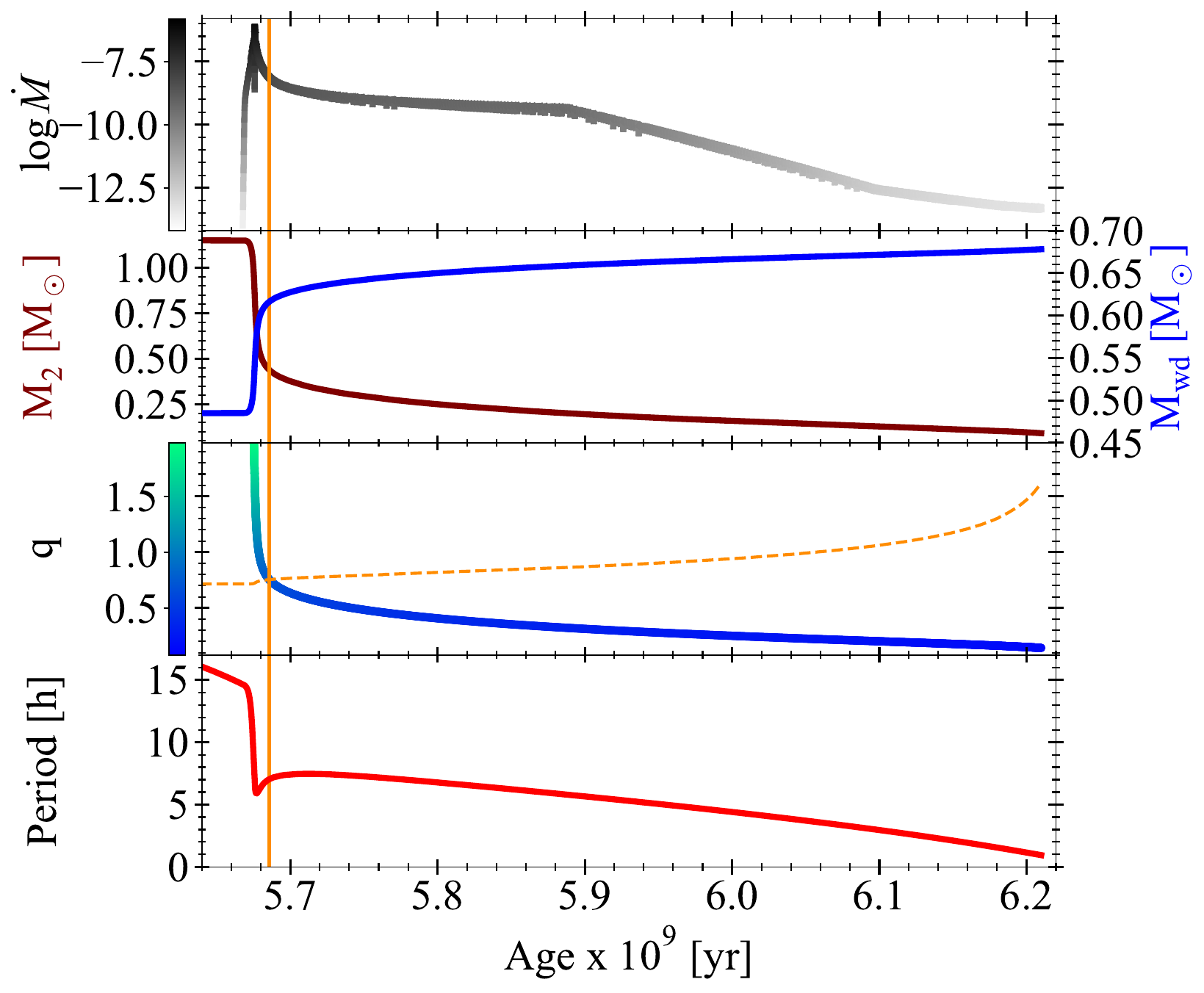} 
    \caption{Evolution of parameters of the best {\sc mesa} model of EI~Psc 
    with the system age. The mass transfer rate $\dot{M}$ is given in 
    $M_\sun\ year^{-1}$ unit. The  dashed orange line shows the critical 
    mass ratio (see the text). The  solid orange line corresponds to the age 
    of the system, after which the mass transfer becomes stable.}
    \label{F:MESAEvolv}  
\end{figure}

\begin{figure*}[t]
    \centering   
    \includegraphics{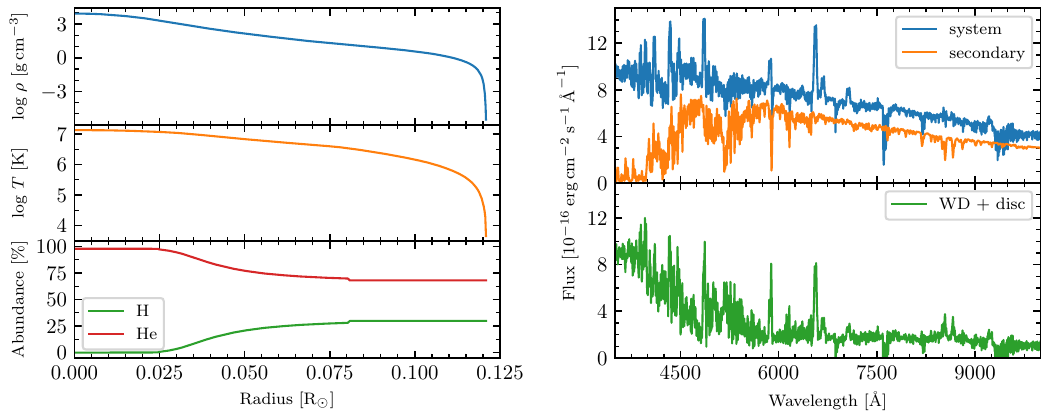} 
    \caption{Left panel: Log of density, log of the temperature, and 
    abundance of H and He in the secondary of EI~Psc following {\sc mesa} model. 
    Right panel: Calculated spectrum of the secondary compared with observed 
    one of EI~Psc. }
    \label{F:EvolvedSecondary}  
\end{figure*}

\section{System evolution}
\label{S:Evol}

As predicted by \citet{1985SvAL...11...52T}, a donor in CV progenitors with 
$1.0 \lesssim M_2/{\rm M}_\odot \lesssim 1.5$ can exhaust a significant fraction 
of the hydrogen in their cores before Roche lobe overflow (RLOF). At a low central 
hydrogen abundance\footnote{The mass-fraction of hydrogen and helium are called 
“X” and “Y” respectively. The symbol “Z” is the mass fraction of all other 
elements heavier than helium. } of the donor ($X_c \lesssim 0.1$),  CVs evolve 
into ultra-compact binaries with $P_{\rm orb} < 80$ min.
Following the prediction and to investigate the 
origin of  EI~Psc, we calculated binary evolution models using the 
{\sc mesa}\footnote{MESA, version 24.08.1} code \citep{2011ApJS..192....3P, 
2013ApJS..208....4P, 2015ApJS..220...15P, 2018ApJS..234...34P, 
2019ApJS..243...10P} installed on the multi-core multiprocessor server of the 
Laboratory of Astrophysics of the Kazakh National 
University\footnote{\url{https://astro.kaznu.info/Computingcluster.html}}.

The binary capabilities of {\sc mesa} are described in detail by 
\citet{2015ApJS..220...15P}. We follow the mass-losing star with detailed 
calculations and WD as a point mass. The mass transfer rate during Roche lobe 
overflow is determined following the prescription of \citet{1990A&A...236..385K}. 
We accepted a non-conservative mass transfer with a loss of $\sim$60 -- 100\% of 
mass transferred from the donor. Mass loss during nova eruptions is the most 
likely mechanism \citep{2016MNRAS.455L..16S, 2020NatAs...4..886H,  
2022ApJ...938...31S} of non-conservative mass transfer and angular momentum loss; 
however, some others were proposed, such as wind-lost or binary-driven mass loss, 
mass-lost from the outer Lagrange points $L_2, L_3$, and formation of circumbinary 
disc, or all together with different contributions during distinct epochs of 
evolution of the system were suggested to explain the discrepancies between 
observations and theoretical predictions of CVs (see \citet{2024ApJ...977...34T} 
and references therein). Nevertheless, we note that there exist some 
contradictions in the predictions of nova-ejection models. For example, there are 
various studies that support the lack of strong growth or permanence of the white 
dwarf mass through accretion \citep{2005ApJ...623..398Y,2010Natur.463..924G,  
2007MNRAS.374.1449E, 2022MNRAS.510.6110P}, however, \citet{2020ApJ...895...70S, 
2024ApJ...962..191S} predict an increase in the masses of WDs in CVs during their
evolution.  In the case of EI~Psc, a low or absent increase in the mass of the
white dwarf is needed to reach the observed mass ratio observed in the system, 
which is relatively high for its orbital period.

In our {\sc mesa} probe, the initial orbital period of the system was selected as 
1.0 -- 1.5 days, with initial donor and WD masses of 1.0 -- 1.5 M$_\odot$ and 0.40 
-- 1.00 M$_\odot$, respectively. We also vary the mass loss (60 -- 100\%) in the 
system by changing a fraction of the mass lost from the vicinity of the accretor 
($\beta$ parameter in the binary mass transfer control of {\sc mesa}).  The last 
was used as a parameter which represents all possible mechanisms~\footnote{ We 
note that the modelling does not explicitly include several 
difficult-to-incorporate effects, such as the influence of nova eruptions 
on the evolution of the secondary star, short-term episodic mass loss from the 
system, and irradiation of the secondary. However, we have assumed that the 
cumulative impact of these processes is effectively captured through the 
time-averaged parameter $\beta$. This assumption is justified by the variable 
time-step used in our modelling, which is controlled by the parameter 
$varcontrol\_target$ = 10$^{-3}$. The resulting time resolution is of the order of 
$\sim10^4$ years during phases of high mass transfer and $\sim10^5$-$10^6$ years 
during stable mass transfer, significantly longer than the duration of  nova 
events (typically lasting from a few days to a few weeks) and comparable to or 
greater than typical nova recurrence timescales. 
The standard {\sc mesa} code was also modified to exclude non-physical 
"resonance" in system evolution, which sometimes occurred due to poor temporal 
resolution in the evolution model. This was achieved by introducing mechanisms to 
limit abrupt changes in key parameters such as radius and luminosity between the 
calculation steps. Corrections were also made to the computation of mass transfer 
rates in binary systems by applying adaptive constraints to avoid unstable spikes. 
The calculation time step was refined to improve the accuracy of tracking rapid 
changes in physical parameters, ensuring a stable and physically realistic 
representation of the system's evolution.} of loss of the accreting material 
and the angular momentum from the system mentioned above. We calculated 
a mesh of models that contains about 600 evolution tracks in total 
The evolution tracks for a selection of {\sc mesa} models are shown in
Fig.~\ref{F:Evmodel} where the distribution of effective temperatures of CVs 
donors vs. orbital periods is also shown by black points (an update of the 
\citet{1998A&A...339..518B} list). The colour of the tracks marks the mass ratio 
of the systems. The gray-scale line around the best model track (the wide solid 
line) represents the full mass transfer rate from the secondary to WD. A part of 
accreting matter is accumulated by WD and the rest leaves from the system.
Systems with WDs $\lesssim 0.6 $M$_\odot$ and non-conservative mass transfer 
produce during its evolution warm secondaries which perfectly explain the sample 
of such systems\footnote{We note that five objects from the sample are located in 
the period gap. Their visibility in this range of periods is probably caused by 
the presence of warm secondaries.} (see Table~\ref{T:WarmSecondary})  marked by 
violet points in the plot. 
Increasing the initial mass of the secondary leads to increased temperature at 
the orbital periods $\approx$1 h, but increasing the accreting mass lost from 
the system provides the opposite effect. The more massive WDs move evolution 
tracks to standard form (dashed lines at Fig.~\ref{F:Evmodel}) while the longer 
initial orbital periods provide the formation of short period $<1$ h systems with 
lower temperatures of the secondaries.
In addition to the non-conservative mass transfer models, the evolution track 
of a "standard" CV with initial $M_{\rm WD}$ = 0.7~M$_\sun$, $M_{2}$ = 
0.5~M$_\sun$ without the loss of accreting material from the system is presented 
by the dashed-point line for comparison. 
The best solution\footnote{The best model solution ($M_{\rm WD}$ = 0.697~\ms, 
$q=0.188$, $\dot{M} = 4.98\times10^{-13}$~\ms\ year$^{-1}$ at current state of 
the system,  the time from the beginning mass transfer to the current state 
$t_{acc}=5.3\times10^{8}$ years) was selected by comparing the orbital period, 
the effective temperature of the secondary, and the mass ratio of the components 
obtained from the evolutionary model with the values derived by light curve 
modelling.} for EI~Psc was found for a system formed from a binary with 
$M_{\rm WD}$ = 0.45~\ms, $M_{\rm donor}$= 1.15~\ms, the initial orbital period 
of 1.17~day and the loss of approximately half of the total mass of the system 
during its evolution or 75\% of the material accreting between components 
($\beta = 0.75$). The peak mass transfer rate reaches 
$\approx$$10^{-6}$~\ms\ yr$^{-1}$ in the beginning (see Fig.~\ref{F:MESAEvolv}), 
after that, it drops to $\approx$$10^{-9}$~\ms\ year$^{-1}$ during 
$\sim$$2.5\times10^8$ years ($P_{orb} \approx 6h$) and late slowly decreases to 
$\approx$$10^{-13}$~\ms\ year$^{-1}$ at the current state. The largest mass 
exchange occurs when the system´s orbital period diminishes from  8 to 6 hours 
and there is a high and unstable mass transfer.
At this time, the most significant loss
of the mass and  the angular momentum in the system occurs.
It is not a surprise because the system still has the mass ratio which 
significantly exceeds the value for steady mass transfer $q_{\rm crit}\approx 
0.65$ (see \citet{1987ApJ...318..794H, 1997MNRAS.291..732T} and references 
therein). The orange dashed line in Fig.~\ref{F:MESAEvolv}, ($q$ vs. age panel)
shows a critical mass-ratio $q_{\rm crit}$, above which mass transfer proceeds 
on a dynamical timescale and a common-envelope (CE) evolution can occur. 
The latter is proposed as the origin of the early post-outburst spectra 
of many novae \citep{2024ApJ...975..191W} and a significant mass loss 
\citep{ 2020ApJ...895...29M}. We also note that high-mass transfer nova-like 
systems with orbital periods within the mentioned range were suggested 
to exhibit an outflow from the accretion disc at the side opposite to the 
secondary \citep{2020MNRAS.497.1475S, 2017MNRAS.470.1960H}. 
At shorter orbital periods (right from the solid orange vertical line in 
Fig.~\ref{F:MESAEvolv}), the value of the mass transfer rate decreases to the 
usual one for CVs, and the mass ratio between components corresponds to the 
maintenance of stable mass transfer. The mass of WD is slowly increasing, while 
the donor mass is decreasing, which is accompanied by a loss of part of the 
accreting material by some of the above-proposed mechanisms. Taking into account 
the loss of the total system mass ($\approx0.2~$\ms) along  time of evolution 
during the stable mass transfer and  an average ejection of a mass caused 
by possible nova eruptions ($\sim$10$^{-5}$--10$^{-4}$\ms, 
\citet{1986ApJ...308..721T, 2005ApJ...623..398Y}, a recurrent time is 
about $\gtrsim$10$^3$-10$^{4}$ years here.

The evolution results in the formation of a close binary system with orbital
periods below 80~min that contains a donor with atmospheres enriched with helium 
up to about 70\%. It is not a surprise because it is known that around the orbital 
period of EI~Psc, there are several systems classified as helium-rich CVs (see 
Table~\ref{T:EIPscnear}) based on a relatively high ratio of \ion{He}{I}/H$\alpha 
\lessapprox 1 $ line intensity. In Fig.\ref{F:EvolvedSecondary},~left, we show 
profiles of the density (top), temperature (middle), and hydrogen and helium 
abundance of a secondary vs. its radius at the moment of the system evolution with 
the corresponding current state of EI~Psc as follows from {\sc mesa} calculation.  

Based on these results, we calculated a spectrum of secondary 
($T_{\rm eff}$ = 4400~K, $\log$ g = 5.38, $X=0.30, Y=0.68, Z=0.02)$  
using {\sc tlusty} and {\sc sysnspec} codes by \citet{2021arXiv210402829H} 
using the {\sc mesa} model for the abundances of some heavy elements (C, N, O, 
Ne, Mg) and adopting solar abundances for the other. In the top panel of 
Fig.~\ref{F:EvolvedSecondary},~right,  we present the flux-calibrated X-shooter 
spectrum together with the calculated one shifted to the flux from the secondary 
from our light curve fitting (see Fig.~\ref{F:LCmod}),  while in the lower panel, 
the difference between them is shown. The modelling spectrum has a flux maximum 
$F_\lambda$ at $\approx 4600$~\AA\ in contrast to $\approx$~6000~\AA\ of a 
zero-age main-sequence star (ZAMS) of similar effective temperature.
The continuum after the peak decreases more rapidly with increasing wavelength 
compared to ZAMS and its slope is the same as that of the X-shooter spectrum. The 
result of subtraction of observed and synthetic donor spectra has a flat continuum 
after $\sim 4600$~\AA.
Below $\sim 4600$~\AA,  the flux of the continuum shows an increase. 
We speculate that it can be caused by the more complicated model of the secondary 
atmosphere and/or by an additional contribution in the blue part of the spectrum 
from a boundary layer between the disc and WD or radiation from L1 and/or stream flux.  

The number of systems with warm donors
must be significantly lower than the rest of the CVs. Sun-like stars 
($\sim$ 1 \ms) are no more than $\lesssim$8\% of the total Galactic 
stellar population \citep[and reference therein]{2015MNRAS.451..149J} 
and their number will be significantly lower in the progenitors of CVs. 

\begin{table*}[]
    \centering
    \caption{Updated list of CVs with anomalously warm secondary according to 
    \cite{2021PASJ...73.1209W} and \cite{2021MNRAS.505.2051E} with orbital periods 
    $\leq$ 5~hours.}
    \begin{tabular}{lccccr}
    \hline\noalign{\smallskip} 
     System   & Period   & $G$ &  Distance  &  Spec. Type   &     Ref. \\
              & [h]    &  [mag]  &  [pc]  &   Secondary   &         \\
    \hline \hline \noalign{\smallskip}    
\object{KSP-OT-201712a} &0.97  & 17.7   &  1400(130)  & K4.5$\pm$0.5   & 1    \\
\object{V485 Cen}       &  0.98     & 17.9 & 331(7) &    K8 &   2  \\
 \object{EI~Psc}   &  1.07  & 	15.9  & 149(1) & K4 $\pm$ 2 &  3      \\ 
\object{QZ~Ser}         & 2.00  & 	17.3 &  320 &K4 $\pm$ 2 &   4  \\ 
\object{CzeV404~Her}$^*$   & 2.35  &   15.9 &  340 &  K7$\pm1$ & 5\\
\object{V1239 Her} & 2.40   &  17.8  & 440±30  & M0$\pm$0.5 &  6   \\
\object{SDSS J001153.08-064739.2} & 2.41  & 17.2 & 484 & K4$\pm3$   & 7 \\
\object{CSS J134052.0+151341}  & 2.45  & 18.0 &  790 ? & K4 $\pm$ 2 & 8     \\
\object{ASAS-SN 18aan}  & 3.58& 16.8  & 675(+32/-29) & G9      & 9   \\
\object{ASAS-SN 13cl}$^*$   & 4.86  & 17.7 & -- & K4 $\pm$ 2,   & 10  \\   
\object{V1460 Her}$^*$ & 4.99 & 15.0 & 263 & K5,   & 11  \\
\object{ASAS-SN 15cm}  & 5.00 & 18.7 & -- & K2.5 $\pm$ 2.5 &  12   \\
    \hline
    \end{tabular}
    \label{T:WarmSecondary}
 \medskip
 
References: 1 - \cite{2024ApJ...964..186L}, 
           2 - \cite{2004ApJ...604..817P}, 
           3 - this paper,
           4 - \cite{2002PASP..114.1117T}, 
           5 - \cite{2021A&A...652A..49K}, \\
           6 - \cite{2006MNRAS.371.1435L}, 
           7 - \cite{2014ApJ...790...28R},
           8 - \cite{2013PASP..125..506T}, 
           9 - \cite{2021PASJ...73.1209W}, \\
           10 - \cite{2015PASP..127..351T},  
           11 - \cite{2020MNRAS.499..149A}, 
           12 - \cite{2016AJ....152..226T}.
           $^*$ - eclipsing system 
\end{table*}

\begin{table*}[]
    \centering
    \caption{List of dwarf novae CVs with periods between 0.92 and 1.11 hours. }
    \begin{tabular}{lcccccr}
 \hline\noalign{\smallskip} 
    Object      &        &   $G$   & Period & Distance & Type  & Ref. \\
                &        &  [mag]  & [h]    & [pc]     &       &     \\ 
\hline\hline\noalign{\smallskip} 	 	 	 	 
	\object{CRTS J111126.9+571239}	&	&	19.2 &	0.9228 & 558(24) & helium-rich CV  & 1 \\	 	 	 
	\object{KIC 9778689} 	&	         &      	20.0 &	0.9426 & - & CV?  &  2\\	  	 	 
	\object{V485 Cen}	&	       &	17.9 &	0.9839 & 331(7) & helium-rich CV &  3 \\   	 	 
 \object{V418 Ser}     &	                            & 	20.2 &	1.0560 &  1035(300) & helium-rich CV & 4 \\                   \\  
 \object{EI~Psc}    &   &	15.9 &	1.0696 & 149(1) & CV & 5 \\ 	 	 	 
	\object{CRTS J174033+414756} &	& 19.6 &	1.0812 & 	430(70) & CV? &  6 \\  
	\object{CRTS J112253.3-111037} &	& 20.5 &	1.0872 & 494(40) & helium-rich CV & 7  \\ 	 	 
	\object{OV Boo}$^*$	 	&    & 	18.1 &	1.11019 & 211(4) & CV  &  8\\ 
    \noalign{\smallskip}  
    \hline \noalign{\smallskip}
    \end{tabular}
    \label{T:EIPscnear}

    \medskip
    References: 
    1 - \citet{2013MNRAS.431..372C}, 
    2 - \citet{2014MNRAS.438..789R}, 
    3 - \citet{1998AAS..128..277M}, 
     4 - \cite{2020MNRAS.496.1243G}, \\
     5 - this paper,  
     6 - \citet{2014AJ....148...63S} ,
     7 -  \cite{2012MNRAS.425.2548B},
     8 - \cite{2021AA...646A.181S}, 
     $^*$ - eclipsing system
    
\end{table*}

\section{Conclusions}
\label{S:Conclus}

Based on the presented analyses, we conclude that: 

\begin{enumerate}

\item  The system contains a relatively low temperature $T_{\rm eff} = 6130$~K, 
$M_{\mathrm WD} = 0.70(4)$~\ms\ white dwarf.  The mass of the secondary is
$M_2=0.13$~\ms. Its temperature $T_2 = 4440$~K is higher than in
the rest of the systems at close orbital periods. The inclination of the system is 
$i= 44\fdg5(7)$, its mass accretion rate is $\approx$ 4$\times$10$^{-13}$ \ms\ year$^{-1}$.

\item The long-term light curve of the system shows outbursts and superoutbursts 
of magnitudes $\approx$ 3 and $\approx$ 4, respectively. The quiescence V-band and 
TESS light curves are double humped and are formed by a combination of radiation 
from the secondary and the hot spot. The variation in the shape of the quiescence 
light curve is related to a change in the contribution of the hot spot that occurs 
over a timescale of months. 

\item The outer disc radius in quiescence is about two times smaller than the 
tidal truncation radius. The radiation from the disc in the continuum in 
quiescence corresponds to a low effective temperature $\sim$1500~K, which means 
that the disc radiates mostly in emission lines and from the hot spot at the 
impact region between the disc and the accretion stream from the secondary.

\item The trailed spectra and Doppler maps of the Balmer lines show the presence 
of doughnut-shaped emission from the accretion disc, a hot spot, and a non-uniform 
extended structure on the side of the disc opposite the hot spot. The \ion{He}{I} 
Doppler maps show structures similar to the ones present in maps of Balmer lines. 
The \ion{He}{II} line is probably formed on the surface of WD or in the boundary 
region between the WD and the disc.

\item Inside of the {\sc mesa} approach we show that systems with warm secondary 
originated from binaries with initial days-scale orbital periods and which 
contained low-mass white dwarfs $M_{\rm WD}\lesssim$ 0.6~M$_\odot$ and relatively 
massive $M_2 \sim 1.1-1.5$~M$_\odot$ main sequence companions. During evolution, 
systems lose about half of the common mass or $\sim$75\% of the accreting material.

\item In the case of EI~Psc, the mass of the WD has increased due to accretion by 
about $0.25$~M$_\odot$ up to the moment. Such a moderate increment and a loss of 
most parts of the accreting material are needed to form the system with the high 
mass ratio $q=0.185$ and the warm  $T_{2}=4440$~K secondary.

\item The object represents a stage of evolution that leads to the formation of 
AM~CVn-type binary systems. The observed number of similar systems must be 
significantly lower in comparison with the usual CVs caused by a lower forming 
rate of its progenitors and a relatively short time for which the system has a hot 
secondary.

\item  Currently, the secondary in EI~Psc is represented by a dense hot helium core object containing about 30\% hydrogen in its atmosphere.

\end{enumerate}

\section{Data availability}

The spectra obtained at GTC and the photometry obtained at OAN SPM and \ond\ observatories are available in electronic form at the CDS via anonymous ftp to \href{http://cdsarc.u-strasbg.fr/}{cdsarc.u-strasbg.fr} (130.79.128.5) or via \href{http://cdsweb.u-strasbg.fr/cgi-bin/qcat?J/A+A/}{http://cdsweb.u-strasbg.fr/cgi-bin/qcat?J/A+A/}. The detailed models of binary stellar evolution with modifications and enhancements to the standard MESA code used for the analysis are available at \href{https://zenodo.org/records/15531340}{https://zenodo.org/records/15531340}.

\begin{acknowledgements}
The work is based on observations carried out at the Gran Telescopio Canarias
     (GTC), installed at the Spanish Observatorio del Roque de los Muchachos of 
     the Instituto de Astrof\'isica de Canarias, in the island of La Palma, 
the Observatorio Astron\'omico Nacional on the Sierra San Pedro M\'artir 
(OAN-SPM), Baja California, M\'exico, Ond\v{r}ejov Observatory, Czech Republic. 
The authors thank the daytime and night support staff at the OAN-SPM for
facilitating and helping to obtain our observations. SVZ acknowledges DGAPA-PAPIIT 
grant IN119323. This research was partially funded by the Committee of Science of the 
Ministry of Science and Higher Education of the Republic of Kazakhstan (Grant No. 
AP19678376). The research of JK and MW was supported by the project
{\sc Cooperatio-Physics} of the Charles University in Prague.
      This paper includes data collected by the {\sc Tess} mission. Funding for
      the {\sc Tess} mission is provided by the NASA Science Mission Directorate. 
Some of the data presented in this paper were obtained from the B. Mikulski 
Archive for Space Telescopes (MAST). Guoshoujing Telescope (the Large Sky Area
Multi-Object Fiber Spectroscopic Telescope LAMOST) is a National Major Scientific 
Project built by the Chinese Academy of Sciences. Funding for the project has been 
provided by the National Development and Reform Commission. LAMOST is operated and 
managed by the National Astronomical Observatories, Chinese Academy of Sciences. 
Based on observations obtained with the Samuel Oschin Telescope 48-inch and the 
60-inch Telescope at the Palomar Observatory as part of the Zwicky Transient 
Facility project. ZTF is supported by the National Science Foundation under Grants 
No. AST-1440341 and AST-2034437 and a collaboration including current partners 
Caltech, IPAC, the Oskar Klein Center at Stockholm University, the University of 
Maryland, University of California, Berkeley, the University of Wisconsin at 
Milwaukee, University of Warwick, Ruhr University, Cornell University, 
Northwestern University, and Drexel University. Operations are conducted by COO, 
IPAC, and UW.
This work has used data from the European Space Agency (ESA) mission
{\it Gaia} (\url{https://www.cosmos.esa.int/gaia}), processed by the Gaia Data 
Processing and Analysis Consortium (DPAC, 
\url{https://www. cosmos.esa.int/web/gaia/dpac/consortium}). 
Funding for the DPAC has been provided by national institutions, 
in particular, the institutions participating in the Gaia Multilateral Agreement.
We acknowledge with thanks the variable star observations from the AAVSO 
International Database contributed by observers worldwide and used in this 
research.
\end{acknowledgements}

\bibliographystyle{aa}
\bibliography{bibliography}
\begin{appendix}
\begin{figure*}[h]
\section{Additional Doppler maps}
  \centering
  \includegraphics[width=0.8\textwidth]{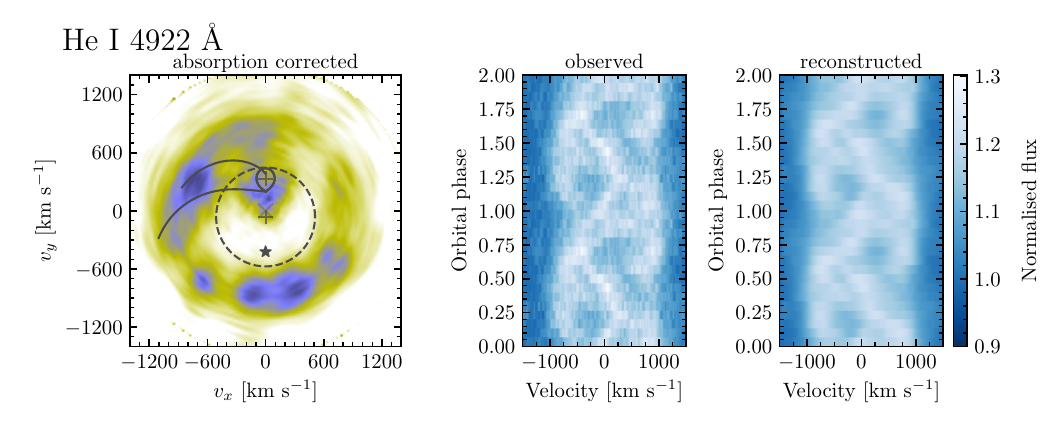}
   \includegraphics[width=0.8\textwidth]{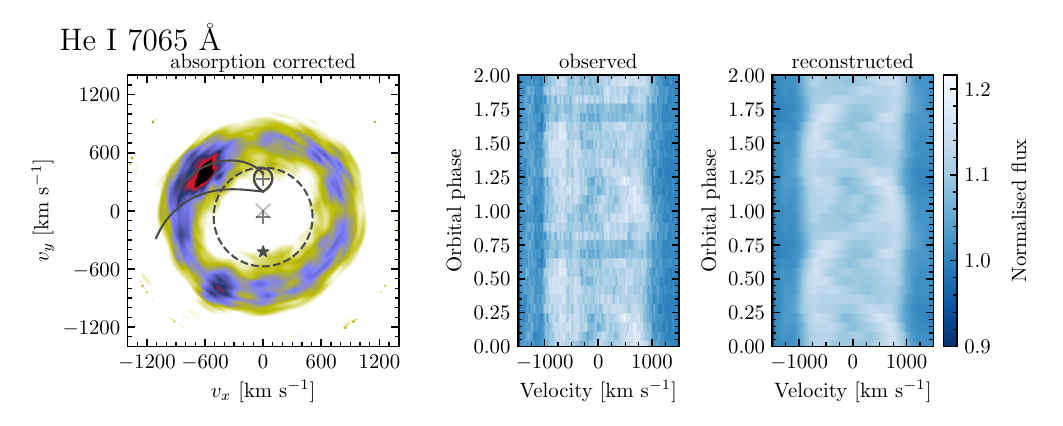}
   \includegraphics[width=0.8\textwidth]{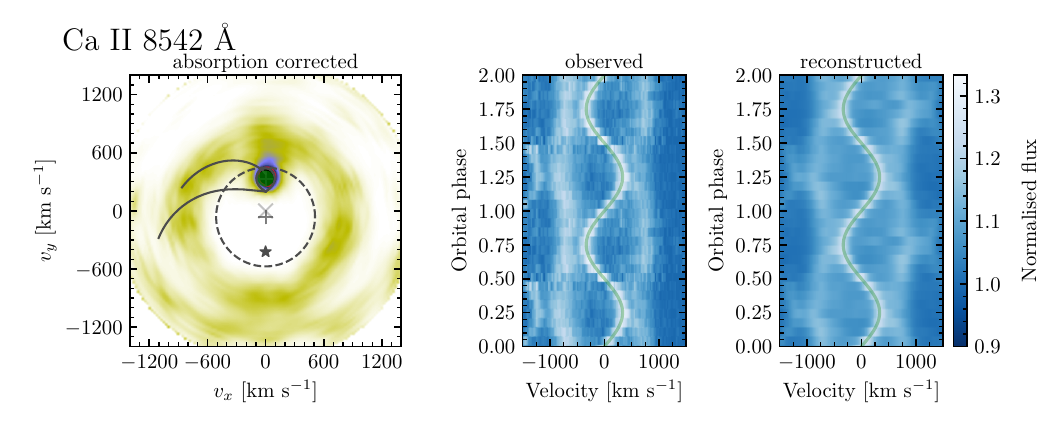} 
  \caption{Absorption-corrected Doppler maps based on \ion{He}{i} and 
  \ion{He}{ii}. See Fig.~\ref{F:DM:AC:H} for a detailed description of shown 
  elements. } 
  \label{AtlasA}
\end{figure*}

\begin{figure*}[t]
\section{Examples of Doppler maps constructed using uncorrected spectra}
   \centering
   \includegraphics[width=0.8\textwidth]{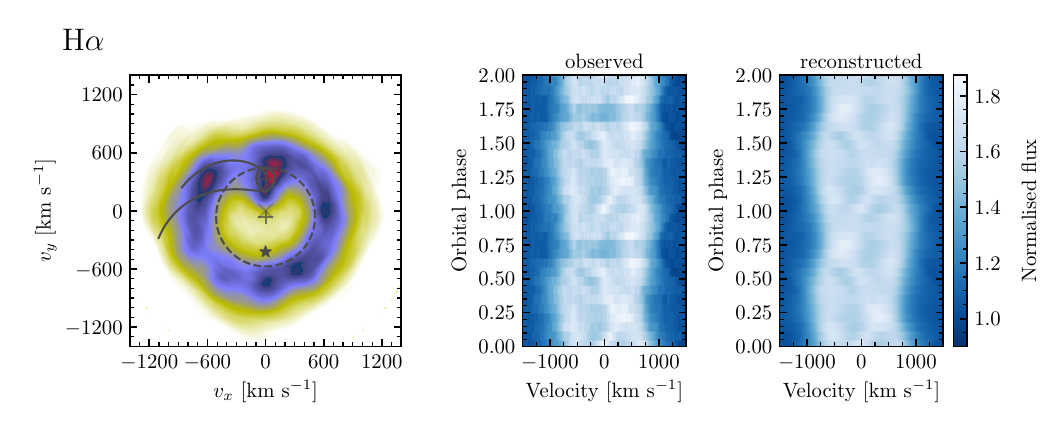}
   \includegraphics[width=0.8\textwidth]{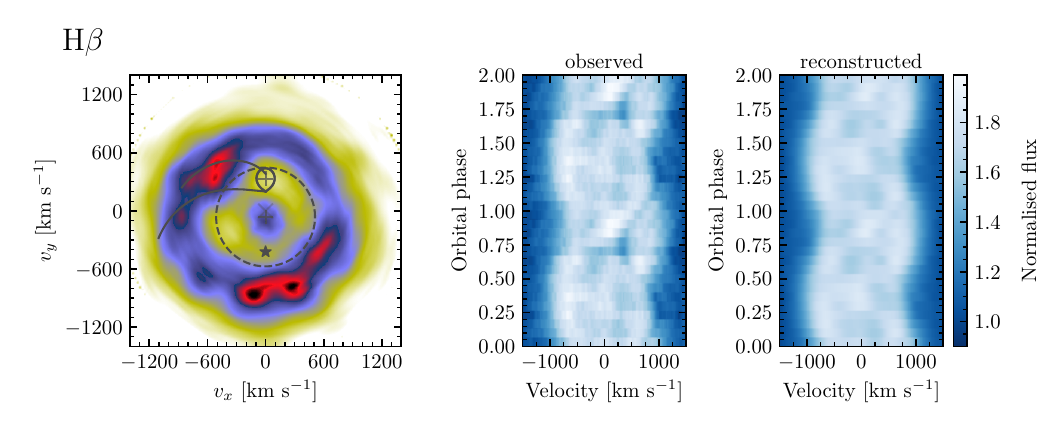}
   \includegraphics[width=0.8\textwidth]{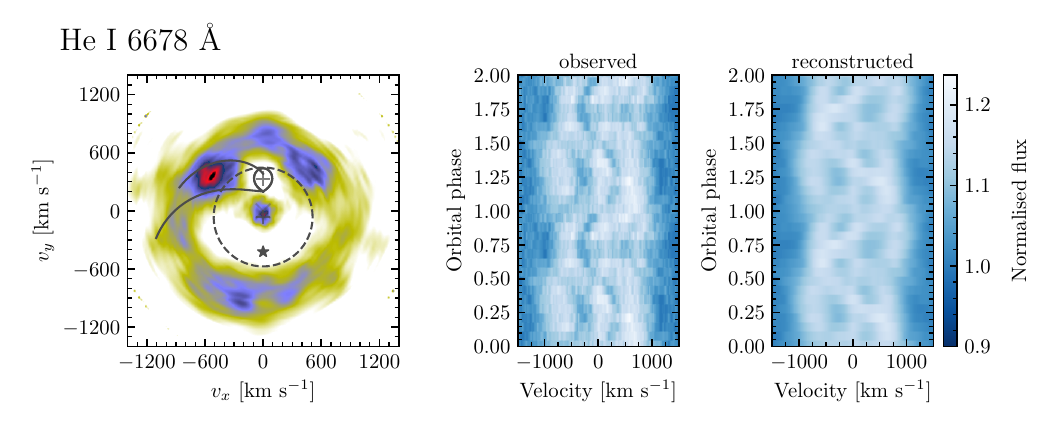}
    \includegraphics[width=0.8\textwidth]{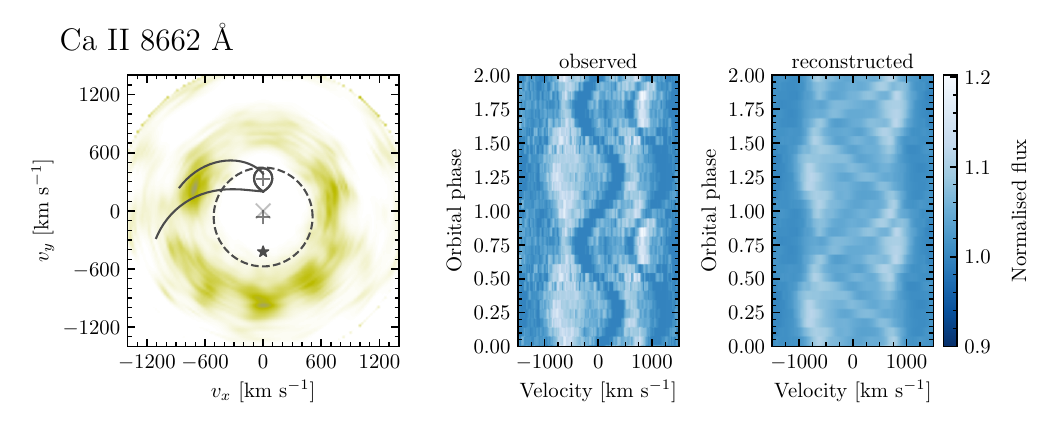}
   \caption{Doppler maps without the absorption correction} 
\label{AtlasB}
\end{figure*}
\end{appendix}
\end{document}